\documentclass[journal]{IEEEtran}

\usepackage{xcolor}

\ifCLASSINFOpdf
\else
\fi
\usepackage[colorlinks=true, linkcolor=blue, citecolor=blue, urlcolor=magenta]{hyperref}

\usepackage{amsmath, amsfonts}
\usepackage{algorithmic}

\usepackage{array}

\usepackage{makecell}
\usepackage{enumitem}
\usepackage{multirow}

\usepackage{algorithm}
\usepackage[caption=false,font=normalsize,labelfont=sf,textfont=sf]{subfig}
\usepackage{textcomp}
\usepackage{stfloats}
\usepackage{url}
\usepackage{verbatim}
\usepackage{graphicx}
\usepackage{epstopdf}
\usepackage{dsfont}
\usepackage{amssymb}
\usepackage{amsthm}

\usepackage{booktabs} 
\usepackage{cite}
\begin{document}
%
\title{Energy-Aware Multi-Exit TinyML for Smart Zero-Energy Devices}
%
%
%

\author{Shahab Jahanbazi,~\IEEEmembership{Student Member,~IEEE,}
        Mateen Ashraf,
        Lieven De Strycker,~\IEEEmembership{Senior Member,~IEEE,}
        Jeroen~Famaey,~\IEEEmembership{Senior Member,~IEEE,}
        Onel~L.~A.~L\'opez,~\IEEEmembership{Senior Member,~IEEE}
\thanks{This work is partially supported in Finland by the Research Council of Finland (Grants 362782 (ECO-LITE), and 369116 (6G Flagship)); and by the European Commission through the Horizon Europe/JU SNS project AMBIENT-6G (Grant 101192113).\\
Shahab Jahanbazi, Mateen Ashraf, and Onel L\'opez are with the Centre for Wireless Communications, University of Oulu, 90014 Oulu, Finland (e-mail: \{Shahab.Jahanbazi; Mateen.Ashraf; onel.alcarazlopez\}@oulu.fi). Lieven De Strycker is with ESAT-DRAMCO, KU Leuven (Ghent), 9000 Ghent, Belgium (e-mail: lieven.destrycker@kuleuven.be). Jeroen Famaey is with IDLab research group, University of Antwerp - imec, 2000 Antwerp, Belgium (e-mail: jeroen.famaey@imec.be).}
}

%
%

\markboth{Journal of \LaTeX\ Class Files,~2026}%
{Shell \MakeLowercase{\textit{et al.}}: Bare Demo of IEEEtran.cls for IEEE Journals}
%



\maketitle

\begin{abstract}
The proliferation of smart and autonomous systems has motivated a shift toward executing intelligence directly on edge devices. This shift becomes particularly challenging for zero-energy devices (ZEDs), where severe constraints on memory, energy availability, and inference accuracy must be addressed simultaneously. In this paper, we present a unified approach to managing these constraints for smart ZEDs. Specifically, we design, train, and deploy a tiny machine learning (TinyML) model for person detection on a ZED. The proposed architecture stores a single model in memory while enabling adaptive inference through multiple exit points, allowing computational effort to scale with input difficulty. As a result, low-energy inference is performed for easy instances, while higher-precision inference is selectively employed for harder cases. This strategy significantly reduces energy consumption without sacrificing detection accuracy. Furthermore, to enhance device autonomy and prevent power failures, we introduce auxiliary energy-aware circuits that dynamically regulate system operation based on available energy. Compared with a state-of-the-art energy-aware single-exit TinyML approach, the proposed method achieves an energy consumption reduction of approximately $29.6\%$. Overall, the proposed framework is appealing for enabling accurate and energy-efficient intelligence on ZED platforms. 
\end{abstract}


%
\IEEEpeerreviewmaketitle

\section{Introduction}
\IEEEPARstart{N}{owadays}, our surroundings have become increasingly populated with small, low-power, and versatile devices. This growing ubiquity has been driven by advancements in hardware development, including energy-efficient sensing technologies, compact microcontroller units (MCUs), and ultra-low-power wireless communication modules~\cite{ref_basic1}. These components collectively enable on-device data processing and communication at milliwatt or even sub-milliwatt power levels~\cite{toward_energy_aware,model_selection}. In particular, the integration of MCUs enables on-device machine learning (ML) capabilities through local data processing and inference. In this context, tiny machine learning (TinyML) has emerged as an effective approach for executing compact ML models directly on resource-constrained MCUs. Such approaches reduce dependence on cloud-based computation while enhancing system responsiveness and data privacy~\cite{ref_basic1,ref_basic2}.

In spite of these advancements in processing capabilities, the need for a reliable and sustainable power source remains a major challenge. Batteries have traditionally been used to supply power to such sensors, but they introduce significant drawbacks, including maintenance overhead, limited operational lifetime, and environmental concerns~\cite{energyaware_sched}. These issues become particularly challenging in large-scale deployments or in remote and hard-to-reach locations. Consequently, there is a growing motivation to develop battery-less smart sensors that rely on ambient energy harvesting rather than traditional batteries. 

Ambient energy harvesters capture readily available energy from the environment, such as light and heat, and convert it into electrical power~\cite{ref_basic1}. However, the harvested energy is typically low in magnitude, variable in intensity, and intermittent in availability. These causes result in producing either alternating current or poorly regulated direct current, depending on the source \cite{onel2025foundation}. To address these challenges, a power management unit (PMU) is employed to rectify, condition, and regulate the input energy and charge a storage element (e.g., a supercapacitor). This storage element provides a stable short-term energy supply while maintaining the device’s compact and environmentally friendly design.

In this paper, a smart zero-energy device (\textit{ZED}) refers to a self-sustaining smart device that operates without an external power supply by harvesting ambient energy and integrating sensing and processing capabilities. Such devices comprise not only an MCU, an energy harvester, and a PMU, but also application-specific modules that enable their targeted functionality (e.g., sensors, actuators, and communication interfaces). For instance, in scenarios where image capture and analysis are essential, components such as cameras and auxiliary sensors are integrated to enable depth or distance measurements \cite{face_detection}. Likewise, when communication with cloud or edge infrastructure is required, the device incorporates suitable transceivers supporting protocols such as LoRa, Bluetooth Low Energy (BLE), or Wi-Fi \cite{face_detection, face_recognition_demo, civil_infrastructure, toward_energy_aware, model_selection}.
This modular and adaptive design philosophy ensures that smart-ZEDs remain energy-efficient, scalable, and well-suited to the diverse requirements of real-world deployments.

The use of energy harvesting in smart-ZEDs raises the following two questions.
\begin{itemize}[leftmargin=1.8em]
  \item \textit{How do smart-ZEDs remain operational in the face of power outages?} Because available ambient energy fluctuates, smart-ZEDs frequently brown out, risking loss of volatile state and partial computations. Avoiding such events requires intermittent-safe operation, which is typically achieved through two energy-aware strategies: (i) checkpointing, involving early brownout detection and proactive storage of critical state in non-volatile memory~\cite{Checkpointing} and (ii) task scheduling, which uses idempotent, task-granular execution to enable correct resumption without redundant computation~\cite{optimal_task_scheduling,energyaware_sched}.
  \item \textit{How do smart-ZEDs deliver the most utility per millijoule (i.e., how to manage scarce energy to maximize the number of completed tasks)}? While preventing failures is essential, it is not sufficient on its own. The objective is to use limited energy resources as efficiently as possible when executing tasks by incorporating timely decision-making.
\end{itemize}
Moreover, because smart-ZEDs are memory-constrained and rely on TinyML inference, maintaining high prediction accuracy is also crucial. This interplay underscores a three-way trade-off among prediction accuracy, energy consumption, and the memory allocated to the compiled model~\cite{ref_basic3}.
In this paper, we address all of these metrics concurrently for smart-ZEDs. To this end, we introduce an energy-aware TinyML model with multiple exit points based on the learning architecture from~\cite{CDL,branchy}, enabling the system to select the earliest exit whose confidence level is sufficient to produce reliable inferences at runtime. This design offers additional flexibility by enabling low-cost inference for easier instances, thereby reducing energy consumption while maintaining overall accuracy. Moreover, the model remains compatible with stringent memory constraints because the auxiliary exit points are implemented using lightweight additional layers that introduce only minimal memory overhead.

\subsection{Related Works}
\begin{table*}
    \centering
    \caption{State-of-the-art methods using an ambient-powered device with either on-device or edge intelligence support.}
    \begin{tabular}{|c|c|c|c|c|c|c|c|}\hline
         & \textbf{Ambient source}  & \textbf{\makecell[c]{On-device\\inference}} & \textbf{\makecell[c]{Multi-exit\\model}} & \textbf{\makecell[c]{TinyML\\ support}} & \textbf{\makecell[c]{Control strategy \\ (Power-gating \\ unused subsystems)}} & \textbf{Energy-awareness} & \textbf{Application} \\\hline
         \cite{underwater_batteryless2} & Acustic energy & $\times$ & $\times$ & $\times$ & \checkmark & $\times$ & Underwater imaging \\\hline
         \cite{civil_infrastructure} & Solar & $\times$ & $\times$ & $\times$ & \checkmark & \checkmark & \makecell[c]{Civil infrastructure \\ monitoring}  \\\hline
         \cite{underwater_batteryless1} & \makecell[c]{Underwater \\ sound} & \checkmark & $\times$ & \makecell[c]{\checkmark \\ (Quantization)} & $\times$ & $\times$ & \makecell[c]{ Underwater animal \\ sounds recognition}  \\\hline
         \cite{face_recognition_demo} & Solar & \checkmark & $\times$ & \makecell[c]{\checkmark \\ (Quantization)} & \checkmark & $\times$  & Face recognition \\\hline
         \cite{batteryless3} & Solar & \checkmark & $\times$ & \makecell[c]{\checkmark \\ (Quantization)} & \checkmark & $\times$ & Face recognition \\\hline
         \cite{face_detection} & Solar  & \checkmark & $\times$ & \makecell[c]{\checkmark \\ (Quantization)} & \checkmark & \makecell[c]{\checkmark \\ Cold start  mechanism } & Face recognition \\\hline
         \cite{toward_energy_aware} & Solar  & \checkmark & $\times$ & \makecell[c]{\checkmark \\ (Quantization)} & \checkmark & \makecell[c]{\checkmark \\ Monitoring \\ capacitor voltage} & Person detection\\\hline
         \cite{model_selection} & Solar & \checkmark & $\times$ & \makecell[c]{\checkmark \\  (Pruning\\+Quantization)} & \checkmark & \makecell[c]{\checkmark \\ Monitoring \\ capacitor voltage} & Gesture detection \\\hline
         This work & Solar & \checkmark & \checkmark & \makecell[c]{\checkmark \\ (Pruning\\+Quantization)} & \checkmark & \makecell[c]{\checkmark \\ Monitoring \\ capacitor voltage} & Person detection \\\hline
    \end{tabular}
    \label{tab:related_works}
\end{table*}
In recent years, substantial research effort has been devoted to the design of ZEDs. Building on these advances, several complementary strategies have been proposed to address the challenges faced and enhance overall performance. Among these, energy-aware strategies have proven particularly effective in enhancing the energy efficiency and sustainability of ZEDs by minimizing energy consumption and extending device lifetime~\cite{onel2025foundation}. In parallel, the deployment of ML models either on-device or within the associated cloud infrastructure improves system adaptability and overall functional performance. We provide a detailed discussion below, in which the features of highly related works are summarized in Table~\ref{tab:related_works}.

\textbf{Energy-aware strategies.}
ZEDs typically require careful hardware–software co-design to handle the variability of harvested energy, thereby reducing the likelihood of power failure during execution~\cite{optimal_task_scheduling,energyaware_sched,face_detection,toward_energy_aware,model_selection}. In the context of hardware-based strategies, researchers have proposed techniques such as duty cycling~\cite{optimal_task_scheduling,energyaware_sched}, dynamic voltage and frequency scaling~\cite{DVFS}, and selective peripheral gating~\cite{face_detection,civil_infrastructure,toward_energy_aware,model_selection}. In practice, such mechanisms monitor available energy and adjust the device’s execution progress accordingly. These methods aim to reduce idle power consumption and to align computation with periods in which sufficient energy is available.
For example,~\cite{civil_infrastructure} introduces a control strategy to disconnect the unused subsystems during idle intervals, achieving a $2.3\times$ reduction in energy per pixel. In terms of software-based strategies, intermittent-execution frameworks have been developed to ensure correctness and forward progress under frequent power failures. Examples of these frameworks include fine-grained checkpointing and task scheduling models tailored to energy-harvesting platforms. More specifically, the former periodically preserves the execution state of a program at carefully selected points, allowing computation to resume correctly after power failures without re-executing completed work~\cite{Checkpointing}. In parallel, the latter strategies may prioritize tasks based on urgency and energy cost, or defer non-critical work when the available energy is limited.~\cite{optimal_task_scheduling}.

\textbf{Smart-ZEDs.}
Smart-ZEDs have recently been considered for different purposes, including object detection and recognition systems~\cite{face_detection,face_recognition_demo,batteryless3,toward_energy_aware,model_selection}, underwater environmental monitoring platforms~\cite{underwater_batteryless1,underwater_batteryless2}, civil infrastructure monitoring~\cite{civil_infrastructure}, and wearable or implantable medical and fitness devices~\cite{apps_wearable}.
Smart-ZEDs incorporate computational intelligence for inference, either via on-device embedded TinyML models (i.e., on-device inference) or via ML models hosted in the cloud (i.e., cloud-offloaded inference)~\cite{tinyml_survey,ondevice_ml}. In the former case, inference is performed locally using on-board resources, while in the latter, the device transmits raw sensory data to the cloud and receives the inferred results in return. Integrating on-device inference capabilities under the strict energy and memory constraints of smart-ZEDs is non-trivial. Indeed, models must fit within a few kilobytes of memory and execute within short, potentially intermittent energy availability cycles. To reconcile the memory requirement, recent works have proposed lightweight model architectures and efficient inference techniques, including pruning~\cite{Pruning}, quantization~\cite{Quantization}, knowledge distillation~\cite{knowledge_distillation}, and the use of hardware-accelerated inference engines when available. Furthermore, prior studies~\cite{toward_energy_aware,model_selection} have integrated energy-aware strategies into these techniques to better align inference performance with the underlying energy constraints.

\textbf{Cloud-offloaded inference.}
One approach to reduce the intelligence requirements for ZEDs is to offload computation to the cloud while the device focuses on data acquisition and communication. In~\cite{civil_infrastructure, underwater_batteryless2}, the authors respectively investigate battery-less wireless imaging systems for smart civil infrastructure and underwater environments, which stream raw image data to the cloud. Despite the cloud's computational gains, overall energy consumption may not decrease if wireless communication processes dominate energy consumption. For instance, as shown in~\cite{civil_infrastructure}, transmission energy consumption accounts for approximately $83\%$ of the total energy consumption, which highlights a fundamental limitation of architectures that rely heavily on raw-data transmission. A complementary perspective is provided by~\cite{underwater_batteryless1}, which demonstrates a battery-less underwater imaging platform that utilizes ambient energy to enable lightweight on-board inference. More specifically, the device classifies four marine mammal species under extreme power constraints and transmits only on-board inference results, reducing energy consumption by $30.19\%$ compared to sending raw data.
Taken together, these works indicate that in smart-ZEDs, communication rather than computation is often the principal bottleneck and suggest shifting more intelligence toward the device, thereby reducing the frequency and volume of data transmissions.

\textbf{Local inference.}
To mitigate the cost of communication and reduce dependence on cloud connectivity, a series of works investigate running TinyML models on ZEDs \cite{underwater_batteryless1, face_recognition_demo, batteryless3, face_detection}. In \cite{face_recognition_demo, face_detection}, the authors propose ARM Cortex-M4-based sensor nodes that execute full convolutional neural network (CNN) inference for face recognition and detection, and transmit only the classification results over LoRa instead of raw images. The work in \cite{batteryless3} introduces a compact, solar-powered face recognition system that combines an integrated camera with an ultra–low-power ML system-on-chip (SoC), sustaining indoor operation while running a six-layer CNN.
Collectively, these works demonstrate the feasibility of local inference on smart-ZEDs and motivate further research into TinyML architectures and runtime mechanisms that can better exploit scarce and intermittent energy resources.

\textbf{Adaptive selection between cloud-offloaded and local inference.}
A natural evolution from purely cloud-based or purely local execution is therefore a hybrid strategy that dynamically combines local and cloud-offloaded inference~\cite{toward_energy_aware, model_selection}. In its simplest form, the device monitors its current energy and time budgets and, based on this information, either performs inference locally or offloads raw sensory data to a cloud service. The cloud performs the inference and returns the result to the device~\cite{toward_energy_aware, model_selection}. Building on this idea, \cite{model_selection} increases on-device autonomy by deploying multiple models with different cost–accuracy profiles and casting the per-inference decision as an energy-aware optimization problem. Specifically, the device selects the most accurate model that can be supported by its current energy reserves by solving a mixed-integer linear programming (MILP) problem before each inference.

Although~\cite{model_selection} proposes a highly adaptive smart-ZED that increases the accuracy and the number of completed tasks, it introduces three drawbacks: (i) the need to store multiple models on a memory-constrained device, (ii) additional energy overhead to solve the MILP problem before each inference, and (iii) the risk of using a high-accuracy model for easy instances which leads to unnecessary energy consumption. It can therefore be inferred that, although energy availability and inference accuracy constitute the primary design constraints for smart ZEDs, memory requirements also represent critical design objectives.
To the best of our knowledge, no prior work has jointly addressed accuracy, memory usage, and energy consumption in the context of smart-ZEDs. These gaps motivate the design of the smart-ZED based on deploying multi-exit TinyML models proposed in this work.

\subsection{Contributions}
In this paper, we present the design and implementation of a compact, solar-powered smart-ZED that operates in an energy-aware manner for person detection. Our system combines a multi-exit TinyML model, energy-aware execution policies, and dedicated support circuits, and is fully deployed on a severely memory- and energy-constrained microcontroller platform. The main contributions of this paper are summarized as follows:
\begin{itemize}
    \item We introduce and benchmark the first energy-aware multi-exit TinyML approach, showing that multi-exit inference consistently outperforms common single-exit models under comparable resource and energy constraints.
    \item We build, train, and deploy a multi-exit TinyML model on ZED's MCU. To fit the model within the tight memory budget of the smart-ZED, we apply pruning and quantization sequentially and successfully compile the resulting network on the memory-constrained platform. We further analyze how model hyperparameters affect both the overall accuracy and the distribution of exited instances across the different exit points.
    \item We design the supporting circuits required to realize energy-aware operation, enabling the device to adapt its inference behavior to the available stored energy. These circuits interface with the MCU to provide the information needed by the proposed execution policies. We demonstrate that the proposed circuits achieve lower energy consumption compared to existing comparable circuit designs.
    \item We develop a smart-ZED prototype that implements the proposed approach and supports on-device person detection. This prototype demonstrates the practical feasibility of jointly optimizing accuracy, memory footprint, and energy consumption under realistic operating conditions.
\end{itemize}

This paper is organized as follows. In Section~\ref{sec:system-model}, we present the system model and associated assumptions. Section~\ref{sec:adap_multi} details the multi-exit tinyML and discusses the relevant design considerations. In Section~\ref{sec:energy_aware}, we introduce the energy-aware operation and the related concepts. Section~\ref{sec:prototype_sec} concludes the paper by describing the prototype implementation and reporting the obtained results. Finally, the conclusions are provided in Section~\ref{sec:conclu}.

\section{System Model and Architecture}
\label{sec:system-model}

\begin{figure}[!t]
  \centering
  \includegraphics[width=2.6in]{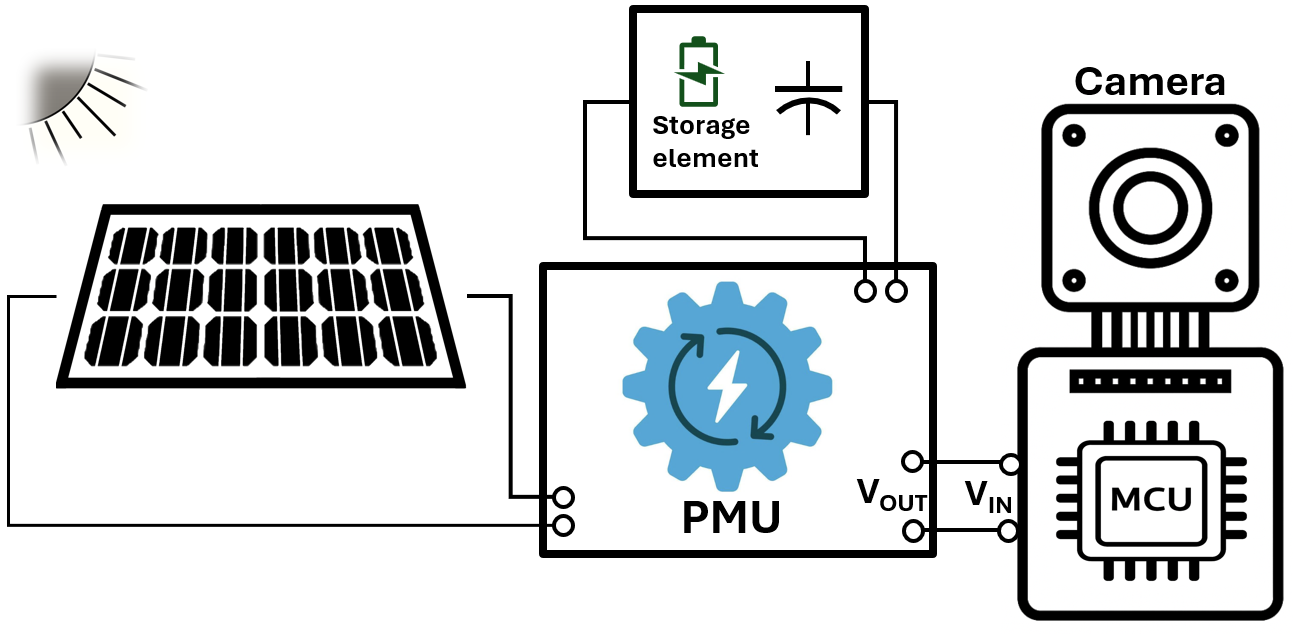}
  \caption{Block diagram of the Smart-ZED architecture: ambient light is harvested by a solar cell and conditioned by a PMU to supply an MCU and to charge the storage capacitor. The MCU orchestrates image capture, on-device inference, and result indication under strict energy and memory budgets.}
  \label{fig:system_model}
\end{figure}

Consider a smart-ZED that must locally capture an image, perform on-device person detection (``person'' vs ``no-person''), and indicate the result, all under tight energy and memory budgets. Fig.~\ref{fig:system_model} summarizes the device architecture, which comprises a solar cell that collects ambient light, a PMU that rectifies, boosts, and regulates the harvested energy, while supplying an MCU and charging a storage capacitor. The MCU draws power from the PMU's output to enable: (i) a low-power scheduler, in coordination with the capacitor voltage to meet the associated constraints; (ii) image acquisition using a camera module; (iii) a pre-processing step sized according to the target model and memory footprint; and (iv) a person-detection model execution and result output.

We utilize a multi-exit model inspired by~\cite{CDL, branchy} to preserve inference accuracy while reducing energy consumption. Specifically, images assigned to an ``easy-instance'' branch are classified using fewer computations, thereby reducing energy consumption. For more challenging inputs, the model proceeds to deeper exits, maintaining accuracy for hard instances. In other words, inference can be interpreted as an adaptive operation. To clarify this, we present the following illustrative comparison. We defer the formal treatment of the model, constraints, and design trade-offs to Section~\ref{sec:adap_multi}.

\subsection{Standard vs. Multi-Exit Inference}
Consider a baseline model $M'$ deployed on the device, occupying $\xi$~KB of dedicated memory and producing a single output with average accuracy $\mathrm{Acc}$ and fixed inference energy~$e$ per input. We compare $M'$ against a two-exit model $M$ that shares the same backbone and introduces an additional early-exit head.

The dedicated memory footprint of $M$ is $\xi+\epsilon$, where $\epsilon \ll \xi$ accounts for the additional parameters and control logic associated with the early-exit head. Thus, there is a memory overhead for supporting adaptive inference but small relative to the backbone model.
Let $e_i, \forall i \in \{1,2\}$ denote the energy consumed when inference exits at the $i$-th exit point of $M$, where $e_1\leq e_2$. In contrast to model $M'$, the energy consumption corresponding to the inference in $M$ is input-dependent: easy instances may exit early and consume~$e_1$, while harder instances require full execution and consume~$e_2$. Following this, the worst-case energy cost of $M$ introduced by inference at the $2$-nd exit point satisfies $e \leq e_2\leq e+\delta$ where $\delta$ is a small overhead induced by the additional exit logic. 
Thus, the worst-case energy cost of $M$ is only marginally higher than that of $M'$. In essence, the average energy consumption lies between $e_1$ and $e_2+\delta$, approaching one or the other according to the frequency of exiting through the $1$-st or $2$-nd exit points, respectively.
Furthermore, let $\mathrm{Acc}_i, \forall i \in \{1,2\}$ denote the average inference accuracy when exiting at the $i$-th exit point of $M$, where $\mathrm{Acc}_1\leq \mathrm{Acc}_2$. By selecting exit points adaptively based on confidence or resource availability, the system can preserve accuracy by allowing hard instances to proceed to later exits while terminating easy instances early. As a result, the overall accuracy of $M$ can approach that of the standard model $M'$, while benefiting from reduced energy consumption on a large fraction of inputs.
All in all, under tight memory and energy constraints, multi-exit inference provides a favorable trade-off: a small increase in memory footprint and bounded worst-case energy overhead enable adaptive energy consumption while preserving inference accuracy.
In the following, we formalize the operating pipeline and assumptions used throughout the paper.

\subsection{Operation Pipeline}
\label{sec:pipeline}
During operation, the device periodically executes the following cycle:
\begin{enumerate}[label=P\arabic*. , leftmargin=2.2em]
  \item Capturing a photo: the MCU powers the camera module and acquires a single image.
  \item Pre-processing: the image is resized, normalized, and quantized to the model’s input format.
  \item Inference: the person-detection network runs on the MCU. Concretely, detections are post-processed to produce a binary decision.
  \item Indication: the MCU drives the on-board LED to report the outcome and then returns subsystems to a low-power state.
\end{enumerate}
To ensure consistent and reliable operation while preventing power outages, we incorporate additional components into the pipeline to monitor available energy and enable energy-aware operation. The idea is to reduce the impact of energy arrival fluctuations at a negligible energy cost. Furthermore, we rely on power-gating techniques to deactivate unused subsystems, thereby reducing energy consumption in inactive modules. For instance, the camera module is powered only during the P1 stage of the pipeline. These mechanisms are detailed in Section~\ref{sec:power_gating_introducing}.

\subsection{Simplifying Assumptions}
\label{sec:assumptions}
For the sake of simplicity, time is partitioned into fixed-length windows of duration $T$ seconds. To keep the model tractable, we make the following assumptions:

\begin{enumerate}[label=\textbf{A\arabic*}. , leftmargin=2.2em]
  \item \textbf{No external capture aids.}
   The system does not rely on mechanical shutters, PIR sensors, or other external triggers to start the pipeline by capturing a photo.
  \item \textbf{Single pipeline per window.}
  In each window, the device attempts at most one full pipeline (P1–P4). No second capture is initiated within the same window. If residual energy remains, leftover energy carries over to the next window. This is for energy frugality and to avoid redundant frames.
  \item \textbf{No intra-pipeline harvesting waits.}
  Once the full pipeline execution is started at a specific time, stages P1–P4 run contiguously without idle gaps to wait and harvest.
  \item \textbf{Energy sufficiency decided up front.}
  Admitting the execution of the full pipeline within a given window requires measuring the available energy before initiation to verify that sufficient energy exists to complete the pipeline. In other words, the device does not continuously monitor available energy to confirm ongoing compliance with the energy requirement while the pipeline is running.
\end{enumerate}

\subsection{Execution Policy}
\label{sec:timing-energy}
\begin{figure}[!t]
  \centering
  \includegraphics[width=3.6in]{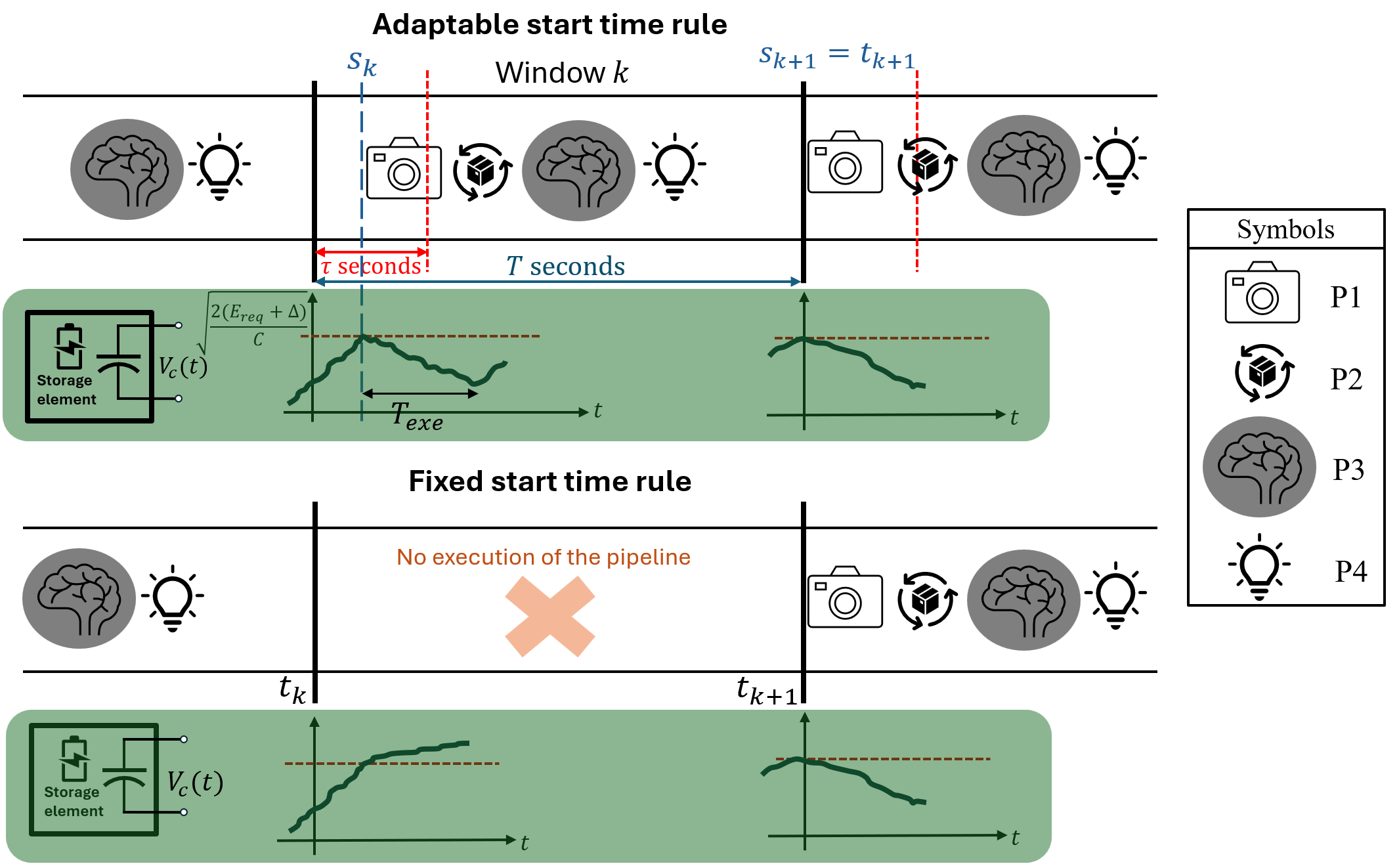}
  \caption{Execution rules considered in this work. Adaptable start time rule allows execution to be deferred until a predefined deadline (top), whereas the fixed start time rule attempts to start executing the full pipeline at the beginning of each window (bottom).}
  \label{fig:execution_rules}
\end{figure}
Let $V_{c}(t)$ denote the capacitor voltage at time $t$. Define the instantaneous energy stored in the buffer as
\begin{equation}
    E_{\mathrm{buf}}(t)=\tfrac{1}{2}C{V_{c}(t)}^{2} \geq 0
    \label{eq:v_c_basic}
\end{equation}
where $C$ is the capacitance. Let $T_{\mathrm{exe}}$ and $E_{\mathrm{req}}>0$ denote the execution time and the energy required to execute one full pipeline, respectively. Furthermore, let $t_k$ be the start time of the $k$-th window and $s_k$ denote the start time of the pipeline execution in the $k$-th window.
For defining when the full pipeline (P1–P4) starts executing (cf. Fig.~\ref{fig:execution_rules}), the device can attempt to begin execution at $N$ discrete times within a permissible window, up to a pre-defined deadline $\tau$, where $\tau+T_{\mathrm{exe}}<T$. Formally, the execution is started in window $k$ at the earliest time $s_k \in \big\{t_k{+}\frac{i}{N}\tau :i\in\{0,1,\cdots,(N{-}1)\}  \big\}$ for which
$E_{\mathrm{buf}}(s_k) \;\ge\; E_{\mathrm{req}}+\Delta,$
where $\Delta\geq 0$ captures conservatism against measurement error and short-term energy variation. If such an $s_k$ exists, the MCU immediately starts execution at $s_k$, completes exactly one full pipeline in window $k$, and then returns to harvest/idle mode. Otherwise, execution is deferred to the $k{+}1$ window. Using $N=1$ as a special case (cf. Fig.~\ref{fig:execution_rules} (bottom)), execution can start only at the beginning of each window (i.e., $s_k=t_k$), thereby avoiding intra-period scheduling complexity and serving as a conservative baseline. However, the first rule relaxes the $s_k=t_k$ requirement to benefit from continuous energy harvesting.
The related outcomes of the two execution policies are presented in Section~\ref{subsec:policies}.

\section{Adaptive Multi-Exit Model}
\label{sec:adap_multi}
Conventional single-exit neural networks execute a fixed number of layers for every input, leading to an almost constant energy cost regardless of input difficulty or the device’s instantaneous energy availability. To address this limitation, prior work proposes adaptive inference through model selection, where one model is chosen from a set of deployed models by solving an MILP problem prior to each inference phase~\cite{model_selection}. While effective, this approach introduces additional control logic and runtime overhead, and increases the on-device memory footprint due to the need to store multiple parameter sets.
In contrast, we propose a multi-exit TinyML model whose inference cost dynamically adapts to both the available energy budget and the confidence of intermediate predictions. Our approach is inspired by BranchyNet~\cite{branchy}, which enables early exiting during inference. A brief review of BranchyNet is provided below.

\subsection{BranchyNet}
The working principle of BranchyNet relies on the fact that inputs to a deep learning network (DLN) vary in difficulty. Specifically, ``easy'' instances can be classified with high confidence, whereas ``hard'' instances remain ambiguous even after extensive training, often yielding misclassifications or low-confidence predictions~\cite{CDL}. Crucially, inferring an instance’s difficulty at runtime is itself nontrivial. As a result, sending every input through the full depth of a DLN can waste computation, add latency, and increase energy consumption, costs that are especially consequential for ZEDs.
To address this inefficiency, conditional deep learning (CDL)~\cite{CDL} adapts computational effort to per-instance difficulty by inserting intermediate exit points comprising lightweight linear classifiers after selected convolutional layers. Specifically, exploiting the hierarchical feature representations learned by CNNs, these exits allow easy inputs to terminate early with minimal computation, while hard inputs proceed deeper for improved accuracy.
However, in CDL, the baseline DLN is trained first and the linear classifiers at prospective exits are then trained iteratively and separately. This decoupled procedure can lengthen training and lead to suboptimal end-to-end performance. The authors in \cite{branchy} address these limitations by introducing the \textit{BranchyNet} architecture, which jointly optimizes the sum of losses from all exits, training the backbone and branches together as a single objective. In this work, we build our model based on BranchyNet architecture.

\begin{figure*}[!t]
\centering
\includegraphics[width=7in]{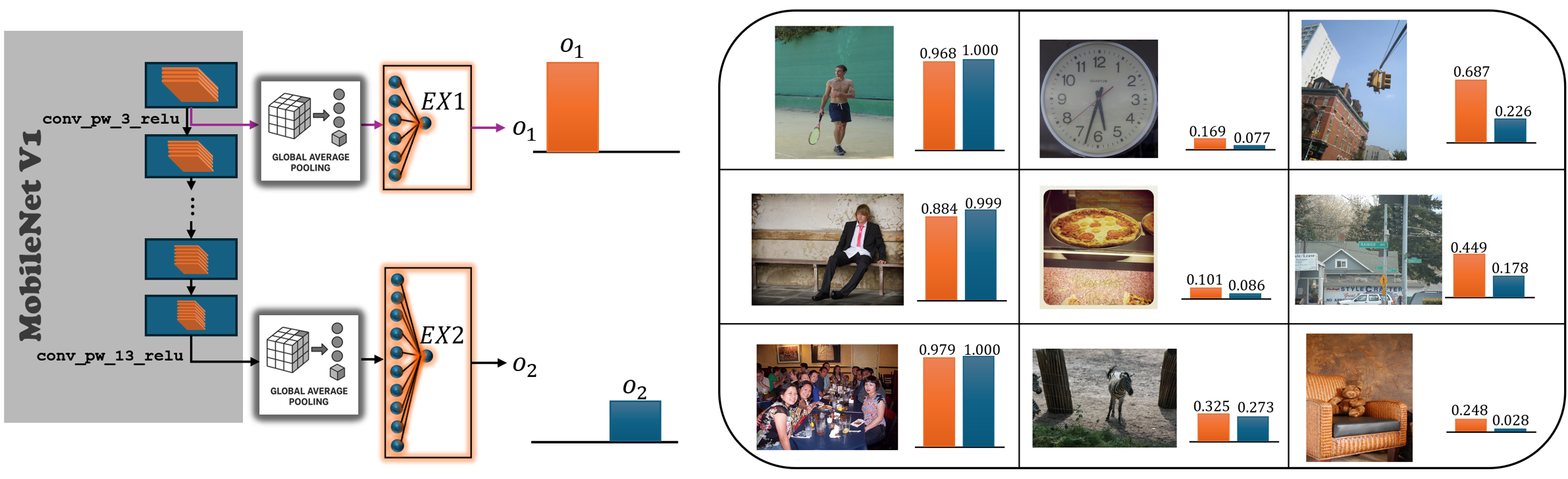}
\caption{Multi-exit network based on MobileNet V1 (left) and the outputs $O_1$ and $O_2$ for some instances by passing them through the trained model (right).}
\label{fig:multi_exit}
\end{figure*}

For the architecture of the employed model, let $M$ consist of a backbone model $M'$ with $J$ ordered exits. To train $M$, we jointly train all its exit points with a weighted sum of losses inspired by BranchyNet \cite{branchy}. Here, the weighted loss function for $N$ input samples can be expressed as
\begin{equation}
    \mathcal{L} = \sum_{j=1}^{J} \lambda_j \,\sum_{i=1}^{N}\mathcal{L}_j(y_i,\hat{y}_{i,j}),
\end{equation}
where $y_i$ denotes the $i$-th input sample, and $\hat{y}_{i,j}$ represents the corresponding output produced at the $j$-th exit of $M$. Meanwhile, $\mathcal{L}_j$ denotes the loss function associated with the $j$-th exit, while the coefficients $\{\lambda_j\}$ serve as weighting factors that balance the contribution of different exit points' performances. 

In this work, model~$M$ is based on MobileNet~V1~\cite{mobilenetv1}, as illustrated in Fig.~\ref{fig:multi_exit} (left). This model belongs to the family of efficient models, designed for energy-constrained environments such as mobile and embedded devices. Additionally, we use the standard MobileNet width multiplier $\alpha$~\cite{mobilenetv1} to effectively reduce the model’s size while maintaining an acceptable level of accuracy.
Moreover, for simplicity, we assume~$M$ has only two branches attached at layers \texttt{conv\_pw\_3\_relu} and \texttt{conv\_pw\_13\_relu}, referred to as \textit{EX1} and \textit{EX2}, respectively. Each branch is connected to $M'$ through a global average pooling (GAP) layer followed by two dense layers. GAP is used to compress the spatial dimensions into one number per channel to achieve a compact and semantically rich feature vector.
It is important to note that the objective of this work focuses on single-class detection. Therefore, each exit’s final layer is designed with a single neuron followed by a sigmoid activation function, leading to a model size, a factor that is particularly critical for ZEDs. Following this, the model outputs are denoted as $O_1\in[0,1]$ and $O_2\in[0,1]$, corresponding to the predicted values at \textit{EX1} and \textit{EX2}, respectively.

\begin{table}[t]
\centering
\caption{Training objective, optimizer, and learning-rate schedule}
\label{tab:adam}
\begin{tabular}{ll}\hline
\textbf{Component} & \textbf{Setting} \\\hline
Framework & TensorFlow/Keras \\
Task & Binary classification \\
Loss & Binary cross-entropy \\
Loss reduction & Mean over mini-batch \\
Optimizer & Adam~\cite{adam} \\
Batch size & $64$ \\
LR schedule & Exponential decay (continuous; no staircase) \\
Initial learning rate & $3\times 10^{-3}$ \\
Decay steps & $1849$ ($\simeq$ one epoch at batch size 64) \\
Decay rate & $0.98$ (i.e., $\eta \leftarrow 0.98\,\eta$ per $1849$ steps) \\
Width multiplier $\alpha$ & $0.25$ \\
Weighting factor $\lambda_1$ & $0.3$\\
Weighting factor $\lambda_2$ & $0.7$\\\hline
\end{tabular}
\end{table}

\subsection{Training setup}
We conduct all experiments using the COCO dataset~\cite{coco2015microsoftcococommonobjects}. This dataset consists of the \texttt{train2017} (118{,}287 images) and \texttt{val2017} (5{,}000 images) splits, which provide bounding-box annotations for 91 object categories. Based on this, we derive a new dataset, the Visual Wake Words (VWW) dataset, by relabeling the images with the labels ``person'' and ``no-person''~\cite{visualwakewordsdataset}. Specifically, each image is assigned a global label $y = 1$ (person) if it contains at least one annotated instance of the COCO \texttt{person} category and $y = 0$ (no-person) otherwise, discarding the bounding-box geometry and retaining only this binary label. This construction yields training and validation sets in which $54.203\%$ and $53.86\%$ of images are labeled as person, respectively, indicating that the dataset is relatively well balanced between the ``person'' and ``no-person'' classes.

We implemented and trained our models in TensorFlow/Keras. Model optimization was performed with the Adam algorithm~\cite{adam}, using Keras' default hyperparameters unless stated otherwise. We adopted a continuous exponential learning rate decay schedule inspired by prior work on lightweight CNNs for mobile deployment~\cite{visualwakewordsdataset}.
The initial learning rate $\eta_0=3\times 10^{-3}$ and decay factor of $0.98$ were selected via a small manual search on a validation set. Specifically, if we index training steps by $t$, the learning rate is $\eta_t = \eta_0\times 0.98^{t/N_{\text{steps/epoch}}}$, where $N_{\text{steps/epoch}}$ denotes the number of mini-batches per epoch, resulting in an effective decay of approximately $0.98$ per epoch. We used a mini-batch size of $64$ and a width multiplier of $\alpha = 0.25$. The training configurations and their corresponding hyperparameters are summarized in Table~\ref{tab:adam}.

\begin{table}[t]
    \caption{Accuracies and \texttt{.h5} size for different width multipliers}
    \centering
    \begin{tabular}{lllll}\hline
         \textbf{Width multiplier}       & \textbf{Network type} &  \textbf{$Acc_1$}   &  \textbf{$Acc_2$}  &  \textbf{Model size}  \\\hline
         \multirow{2}{*}{$0.25$}&  Multi-exit  & $0.7265$ & $0.8309$ & $3{,}185$ KB \\
                                &   Standard   & -        & $0.8425$ & $3{,}159$ KB \\\hline
         \multirow{2}{*}{$0.5$} &  Multi-exit  & $0.7613$ & $0.8657$ & $10{,}508$ KB \\
                                &   Standard   & -        & $0.8914$ & $10{,}469$ KB \\\hline
         \multirow{2}{*}{$0.75$}&  Multi-exit  & $0.7814$ & $0.8934$ & $22{,}435$ KB \\
                                &   Standard   & -        & $0.9267$ & $22{,}381$ KB \\\hline
         \multirow{2}{*}{$1$}   &  Multi-exit  & $0.8030$ & $0.9116$ & $38{,}948$ KB \\ 
                                &   Standard   & -        & $0.9309$ & $38{,}884$ KB \\\hline
    \end{tabular}
    \label{tab:alpha_various}
\end{table}

As illustrated in Fig.~\ref{fig:multi_exit} (right), the model’s outputs vary with input difficulty. In particular, for easy inputs the primary output $O_1$ attains high confidence for the correct label. Hence, the more expensive secondary inference path \textit{EX2} is not required for such inputs. Executing \textit{EX2} incurs higher computational and energy costs on the device. Therefore, a runtime policy should consider both the confidence of $O_1$ and the device’s available energy budget. Based on these factors, the policy can decide whether to accept the prediction produced by $O_1$ or to perform inference using $\textit{EX2}$. This adaptive decision can potentially reduce the average energy consumption while preserving accuracy. A more detailed discussion on this possibility is provided in the next subsection.

Please note that the multi-exit model's training objective is a weighted sum of the losses at its two exit points. Following this, it is worthwhile to measure the resulting degradation in accuracy relative to a standard MobileNet that has a single exit. Table~\ref{tab:alpha_various} reports the classification accuracies at \textit{EX1} and \textit{EX2} for the BranchyNet-based model and different width multipliers, together with the accuracy at the single exit of a conventional MobileNet-based model. The size of the corresponding model file (\texttt{.h5}) is also shown, allowing direct comparison of accuracy and model footprint as the network width varies.

\subsection{Runtime exiting policy}
\label{sec:runtime_policy}
\begin{figure}[!t]
\centering
\includegraphics[width=3.6in]{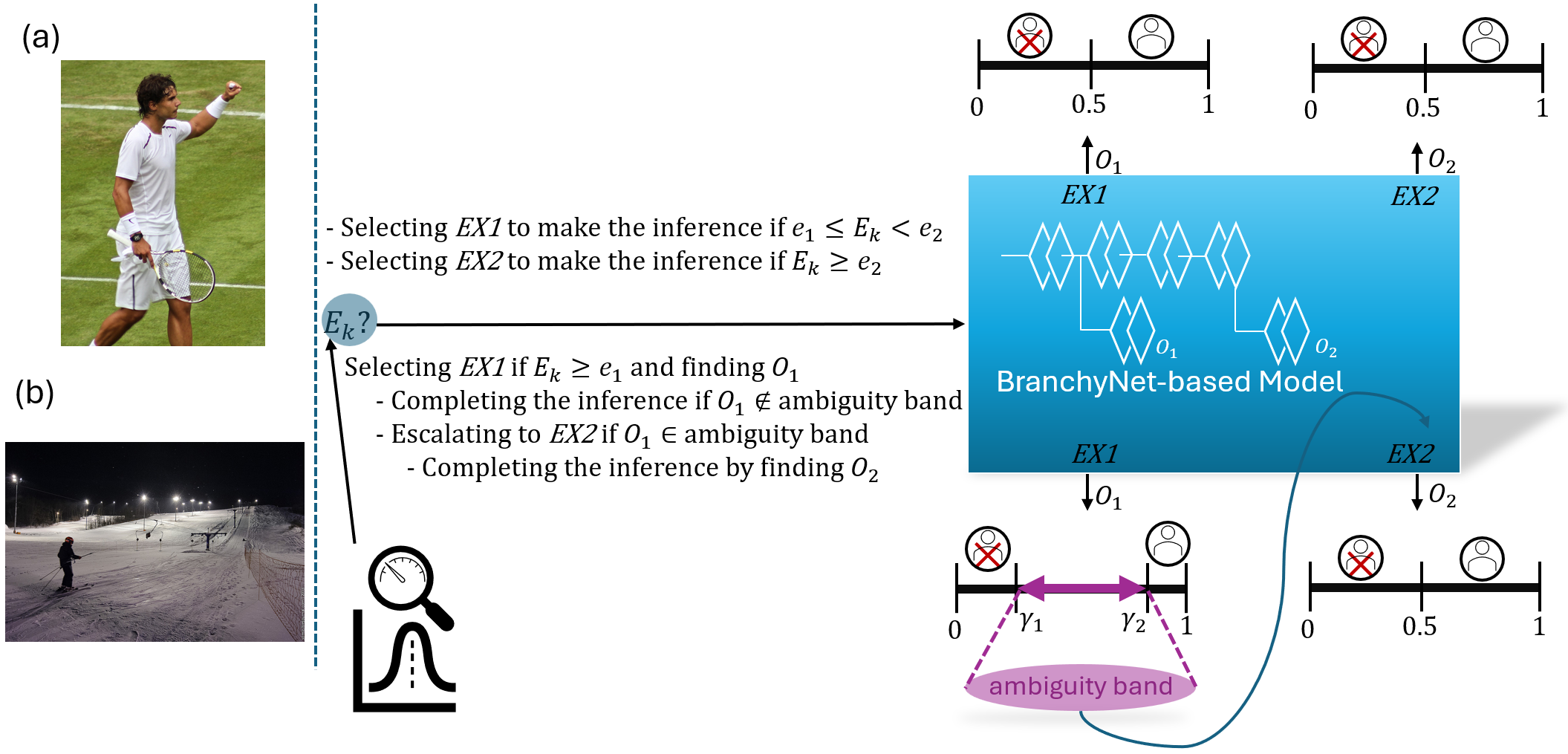}
\caption{Basic policies for runtime selection during inference phase over window $k$: (top details) policy (i) and (bottom details) policy (ii).}
\label{fig:runtime_selection_traditional}
\end{figure}
We restrict our attention to the inference phase of the pipeline (denoted as P3). Let $E_k$ represent the energy available at the beginning of the inference phase within window $k$.
At runtime, the device selects the appropriate inference exit (i.e., \textit{EX1} or \textit{EX2}) based on the observed energy budget $E_k$. This decision is guided by both the available energy and the pre-characterized per-exit performance profiles, given by
$\{(e_1, \mathrm{Acc}_1), (e_2, \mathrm{Acc}_2)\}$.
To balance the trade-off between energy efficiency and prediction reliability, we briefly note two lightweight runtime baselines (see Fig. \ref{fig:runtime_selection_traditional}): policy (i), which precomputes the deepest feasible exit whose estimated energy demand does not exceed the available energy, thereby avoiding the occurrence of power failure during inference under strict energy constraints, and policy (ii), which evaluates \textit{EX1} first and proceeds to \textit{EX2} only when the confidence at \textit{EX1} fails to meet a predefined reliability threshold, thus adjusting inference depth to input difficulty to save energy on easy instances while preserving accuracy on harder ones.

Both policies are computationally lightweight, requiring only integer arithmetic and a few comparisons, and are illustrated in Fig. \ref{fig:runtime_selection_traditional}. In practice, policy (i) guarantees operation within the energy budget but may allocate more energy than necessary for easy instances such as image (a) in Fig. \ref{fig:runtime_selection_traditional}. Conversely, policy (ii) offers input adaptivity and higher energy efficiency and accuracy but may risk triggering a power failure if it escalates to a higher-cost exit (i.e., \textit{EX2}) when the remaining budget is insufficient. As an illustrative example for policy (ii), let $E_k$ be such that $e_1<E_k<e_2$. If \textit{EX1} fails to meet the required confidence threshold, a confidence-gated policy would attempt \textit{EX2}, which may result in a power failure and waste the consumed energy. To mitigate this issue, this work introduces an enhanced policy, derived from Policy (ii), that incorporates an additional measurement point to achieve energy safety and input adaptivity simultaneously just after determining $O_1$.

\subsubsection{Decision rule with a two–exit classifier}
Consider a binary detector (\textit{``person''} vs. \textit{``no–person''}) in which the final layer terminates in a single neuron that outputs a scalar score $O\in[0,1]$. In a conventional single-exit network, a fixed threshold $\gamma$ is used to determine the predicted label. Specifically, it predicts label ``person'' if $O\ge\gamma$ and ``no–person'' otherwise. Meanwhile, with two exits, the model can perform inference at \textit{EX1} with low energy consumption, while leveraging the additional, higher-accuracy exit at \textit{EX2} to improve prediction accuracy when the output confidence at \textit{EX1} is low.

\subsubsection{Uninterrupted and confidence–gated policy}
\label{subsec:runtime_policy}
We now present the runtime selection policy that integrates energy awareness purely to mitigate power-failure risk. The policy is governed by two tunable ambiguity thresholds $(\gamma_1,\gamma_2)$ at \textit{EX1}, where $0\le\gamma_1\le0.5\le\gamma_2\le1$, and two hard feasibility checks based on $e_1$ and $e_2$ with the available energy to prevent energy-induced failure.

In more detail, if $E_k \geq e_1$, it computes $O_1$. If $O_1 \notin (\gamma_1, \gamma_2)$, make the inference at \textit{EX1} as follows
\[
\hat{y}_k=\begin{cases}
\text{``person''}, & O_1\ge \gamma_2,\\
\text{``no–person''}, & O_1\leq\gamma_1.
\end{cases}
\]
In contrast, if $O_1 \in (\gamma_1, \gamma_2)$, the inference escalate to \textit{EX2}. But before that, to avoid a power failure, the device measures the available energy again (i.e., $E_k'$). This second measurement is required because the incoming energy between the two instants is stochastic and cannot be inferred solely from the previous measurement $E_k$ and the consumed energy $e_1$. Although energy measurement incurs a small overhead~\cite{onel2025foundation}, this cost is negligible compared to the risk of incorrect energy estimation. Following this, if $E_k'<e_2-e_1$ (i.e., the escalation is infeasible), the inference will be made at \textit{EX1} as follows
\[
\hat{y}_k=\begin{cases}
\text{(fallback) ``person''},  & 0.5\leq O_1<\gamma_2,\\
\text{(fallback) ``no-person''},  & \gamma_1<O_1<0.5.
\end{cases}
\]
In other words, this branch (i.e., \textit{EX1}) either (a) accepts a confident early prediction, or (b) enforces the energy constraint by prohibiting escalation. Otherwise (i.e., when $E_k'\ge e_2-e_1$ and $O_1\in(\gamma_1,\gamma_2)$), by having the possibility of escalating safely to \textit{EX2}, the device computes $O_2$ and decides with the balanced threshold:
\[
\hat{y}_k=\begin{cases}
\text{``person''}, & O_2\ge 0.5,\\
\text{``no–person''}, & O_2<0.5.
\end{cases}
\]
\begin{figure*}[!t]
\centering
\includegraphics[width=5.5in]{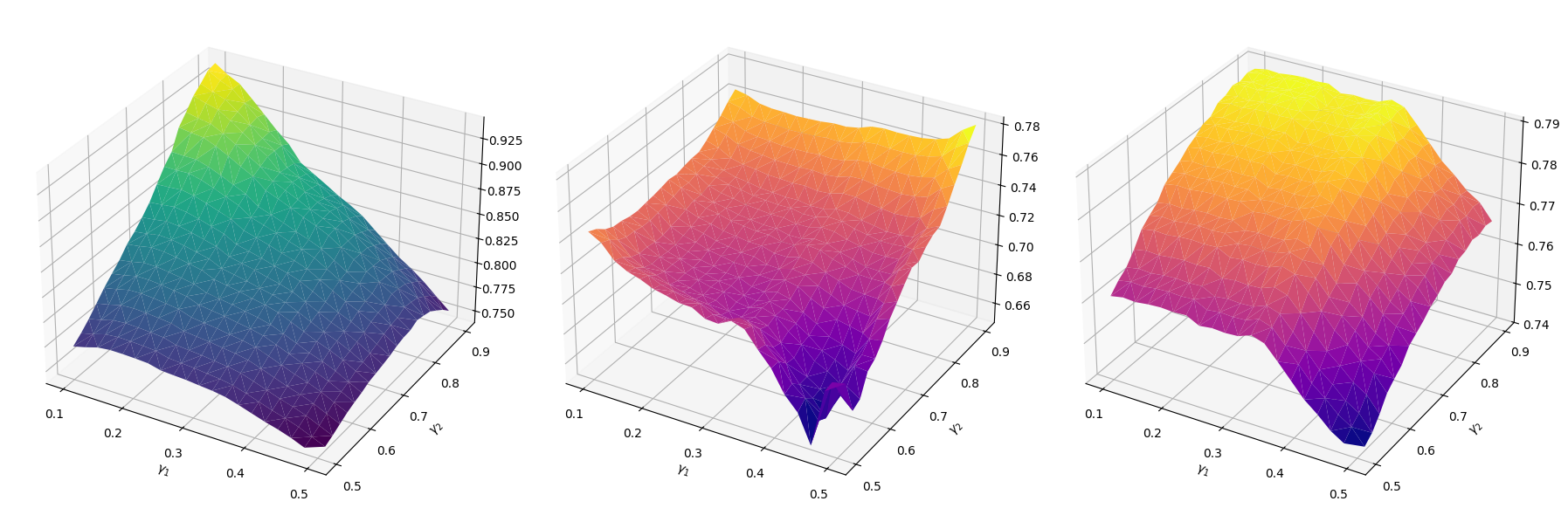}
\caption{Accuracy at \textit{EX1} (left), \textit{EX2} (middle) and total accuracy (right) versus various $\gamma_1$ and $\gamma_2$}
\label{fig:Accuracy_for_all_layers}
\end{figure*}
\begin{figure*}
\centering
\includegraphics[width=4.2in]{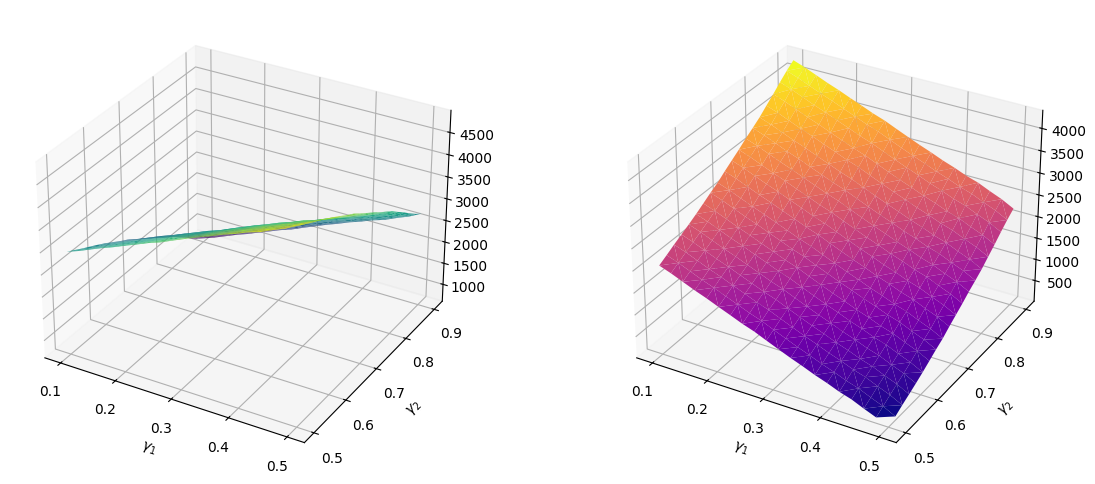}
\caption{Number of validation instances ($5000$ instances) predicted at \textit{EX1} (left) and \textit{EX2} (right) versus various $\gamma_1$ and $\gamma_2$.}
\label{fig:Number_of_exited_samples}
\end{figure*}

We can mention the following properties for the considered policy.

\begin{itemize}[leftmargin=*, itemsep=2pt]
\item \emph{No-power failure guarantee:} Inference at \textit{EX1} occurs only if there is enough energy (i.e., $E_k\ge e_1$). Furthermore, escalation to \textit{EX2} occurs only if $E_k'\ge e_2-e_1$. Both ensure that inference's energy consumption never exceeds the available energy.
\item \emph{Input adaptivity:} Ambiguous or low–confidence cases ($\gamma_1<O_1<\gamma_2$) trigger evaluation by the higher-accuracy exit, while confident cases terminate early to save energy.
\item \emph{Negligible overhead:} The policy requires one additional energy measurement, which incurs only a negligible energy cost (cf. Section~\ref{sec:current_cons}).
\end{itemize}
Please note that (i) setting $\gamma_1=\gamma_2=\gamma$ recovers the classic single–threshold rule (no ambiguity band). (ii) Choosing $\gamma_1<0.5<\gamma_2$ concentrates escalation on uncertain cases, improving accuracy while saving energy on easy inputs. Next, we discuss $(\gamma_1, \gamma_2)$ calibration.

\subsubsection{Threshold-dependent behavior of the inference}
Herein, the first question that may arise is how the parameters $(\gamma_1,\gamma_2)$ should be chosen?
To identify suitable values, we examine the accuracy at \textit{EX1}, \textit{EX2}, and the overall accuracy across a range of $(\gamma_1,\gamma_2)$ settings, as shown in Fig.~\ref{fig:Accuracy_for_all_layers}. In addition, to account for all potential behavioral shifts, Fig.~\ref{fig:Number_of_exited_samples} reports the number of validation instances (out of $5000$) that terminate at \textit{EX1} or \textit{EX2} for each parameter configuration. It is worth noting that as $\gamma_1$ and $\gamma_2$ approach $0.5$, the total accuracy declines due to the combination of low accuracy and a high termination rate at \textit{EX1}. In contrast, when $\gamma_1 \to 0.1$ and $\gamma_2 \to 0.9$, the total accuracy increases because it becomes largely determined by the accuracy at \textit{EX2}, while the number of instances predicted at \textit{EX1} decreases. Consequently, since $e_2>e_1$, the overall energy consumption becomes high in this configuration. The three-way trade-off among accuracy, energy consumption, and the parameters $(\gamma_1, \gamma_2)$ is examined in a practical setting in Section~\ref{sec:tradeoff}.

\subsection{Converting the ML model to a TinyML model}
\label{sec:ML_to_TinyML}
After training and validating the model on the collected dataset, the model is prepared for deployment on the memory-constrained MCU using the TensorFlow Lite (TFLite) library. The TFLite converter exports the trained model to the compact FlatBuffer format and can apply a range of optimizations to reduce computational complexity and memory footprint while preserving acceptable predictive performance. Common optimization approaches include pruning~\cite{Pruning} and quantization~\cite{Quantization}. Pruning eliminates redundant weights or filters to reduce parameter count, whereas quantization reduces the numerical precision of model parameters (e.g., from $32$-bit floating-point to $8$-bit integer) to shrink model size and accelerate inference on integer-friendly hardware. Since the MobileNet V1-based architecture is relatively large for MCU deployment (cf. Table \ref{tab:alpha_various}), we first apply structured pruning followed by pruning-preserving quantization-aware training (PQAT) (with representative-data calibration where required) to obtain a compact \texttt{.tflite} model. This two-stage compression enables the model to meet MCU memory constraints while preserving accuracy for tinyML deployment. The optimized model is then validated to ensure that accuracy remains within acceptable bounds.
In the final deployment stage, the optimized model is embedded into the firmware as a read-only C/C++ array, with all quantization and conversion steps performed offline to minimize runtime overhead on resource-constrained MCUs.

\section{Energy-aware Operation}
\label{sec:energy_aware}
\begin{figure*}[!t]
\centering
\includegraphics[width=5in]{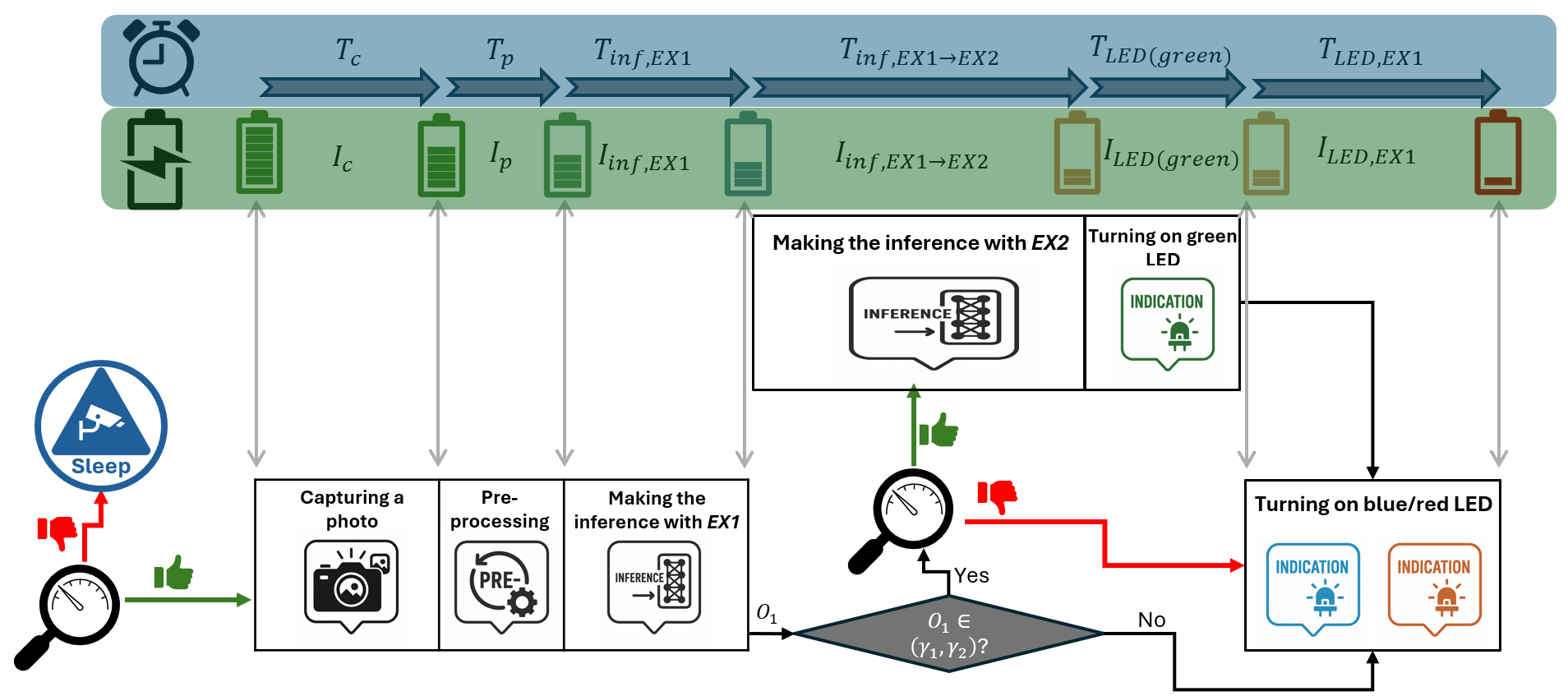}
\caption{Possible operational states of the system performing in an energy-aware manner.}
\label{fig:different_states}
\end{figure*}
In this section, we characterize the energy-aware operation of the device over the complete pipeline P1-P4, as shown in Fig.~\ref{fig:different_states}. Energy harvesting results in time-varying and unpredictable energy arrivals, thus, the energy available at the beginning of an execution period may be insufficient to complete a task, while additional energy may arrive shortly afterward. To account for this temporal variability, the device performs energy measurements at flexible time instants during operation by monitoring the capacitor voltage. These measurements enable the device to initiate computation only when sufficient energy is available to complete the pipeline, thereby preventing power failures and avoiding unnecessary energy waste. Accordingly, we apply the runtime policy described in Section~\ref{subsec:runtime_policy} to the operation pipeline introduced in Section~\ref{sec:pipeline}. More specifically, in each window, the device must first measure the available energy required to complete the process by making an inference at \textit{EX1}. This measurement operation can comprise only one point or include more than one point (see Fig.~\ref{fig:execution_rules}). Furthermore, if the resulting output $O_1$ falls within the ambiguity band, the device must then perform an additional measurement to determine whether completing the inference at \textit{EX2} is feasible. This second measurement also contributes to improving the overall accuracy of the decision. Fig.~\ref{fig:different_states} illustrates the potential states and the locations where the available energy needs to be measured. Note that, to facilitate local observation of the inference outcome, we use the onboard LED as a status indicator. When the inference is executed on \textit{EX1}, the LED turns blue or red to indicate detection of a person or no person, respectively. When the inference is executed on \textit{EX2}, an additional green illumination indicates that the inference has completed on \textit{EX2}.

\subsection{Measurement point and method}
We measure the voltage directly across the energy buffer capacitor, $V_{c}$, via the MCU's analog-to-digital converter (ADC) through a high-impedance divider. Sampling uses a low duty cycle (to limit overhead) and a short moving average (to reduce quantization noise). The divider and ADC are calibrated once, and each sample provides an immediate estimate of the stored energy via~\eqref{eq:v_c_basic}. 

\subsection{Energy consumption model}
We adopt the state-based energy model considered in~\cite{toward_energy_aware, civil_infrastructure}. The MCU and camera module operate at a constant supply voltage of $V_s=3.3$V. Moreover, for each pipeline state $\kappa$ (e.g., \emph{idle}, \emph{capture}, \emph{preprocess}, \emph{inference}, and \emph{indication}), the system draws a state-specific average current $I_\kappa$ for a duration $T_\kappa$ (cf. Fig. \ref{fig:different_states}). Accordingly, the energy consumed in state $\kappa$ at the regulated rail $V_s$ is
\begin{equation}
E_\kappa \;=\; V_s \, T_\kappa \, I_\kappa.
\label{eq:Et}
\end{equation}

To employ the energy-aware operation, under A3 (Section~\ref{sec:assumptions}) and runtime policy mentioned in Section~\ref{subsec:runtime_policy}, a full pipeline (P1–P4) executed with \textit{EX1} requires
\begin{align}
E_{\text{req},\,\textit{EX1}}= V_s\Bigl(
T_c I_c + T_p I_p &+ T_{\text{inf},\textit{EX1}} I_{\text{inf},\textit{EX1}}\nonumber\\
&+ T_{\text{LED},\textit{EX1}} I_{\text{LED},\textit{EX1}}
\Bigr),
\label{eq:EreqExit}
\end{align}
where subscripts ``\(c\)'' and ``\(p\)'' denote the \emph{capture} and \emph{preprocess} states, respectively, while the subscripts ``$\text{inf},\textit{EX1}$'' and ``$\text{LED},\textit{EX1}$'' correspond to the \emph{inference} and the associated on-board LED color (i.e., blue or red) indication when \textit{EX1} is taken. By definition of $e_1$, we have
\begin{equation}
    e_1= V_sT_{\text{inf},\textit{EX1}} I_{\text{inf},\textit{EX1}}.
\end{equation}
Following this, when $O_1$ falls within the ambiguity band, the device needs carry out additional measurement again. Specifically, it needs to determine the remaining energy required to complete the process through \textit{EX2}. For this purpose, we note that this energy comprises only the inference energy for the remaining part of the model given by
\begin{equation}
e_2-e_1=V_sT_{\text{inf},\,\textit{EX1}\to\textit{EX2}}I_{\text{inf},\,\textit{EX1}\to\textit{EX2}},
\end{equation}
and the energy needed to activate the green LED (i.e., $V_sT_{\text{LED(green)}} I_{\text{LED(green)}}$) plus the blue or red color of the LED. Therefore,
\begin{align}
E_{\text{req},\textit{EX1}\to\textit{EX2}}= V_s(&
T_{\text{LED(green)}} I_{\text{LED(green)}} + T_{\text{LED},\textit{EX1}} I_{\text{LED},\textit{EX1}}\nonumber\\
&+T_{\text{inf},\,\textit{EX1}\to\textit{EX2}}I_{\text{inf},\,\textit{EX1}\to\textit{EX2}}).
\end{align}

\subsection{Power-failure-aware scheduling}
A power failure at a specific time occurs if the capacitor voltage at that time falls below \(V_{\text{off}}\) during execution. To prevent this, we compare the pipeline’s required energy to the usable energy in the capacitor. Specifically, let $V_{c}(t)$ be the capacitor voltage at time $t$ in a window in which the device attempts to start executing a pipeline. In addition, let us assume the worst-case final voltage at completion equals \(V_{\text{off}}\). Therefore, if $E_{\text{avail}}(t)$ denotes the usable energy at time~$t$, it can be expressed as
\begin{equation}
E_{\text{avail}}(t) \;=\; \tfrac{1}{2} C \bigl(V_{c}(t)^{2}-V_{\text{off}}^{2}\bigr).
\label{eq:Eavail}
\end{equation}
To guarantee completion with \textit{EX1}, the device requires
\begin{equation}
E_{\text{avail}}(t) \;\ge\; E_{\text{req},\,\textit{EX1}}+\Delta,
\label{eq:guard}
\end{equation}
and in a similar manner, if the completion process is escalated to \textit{EX2} due to the resulting value of $O_1 \in (\gamma_1,\gamma_2)$, an additional measurement is performed to check the feasibility of performing this transfer. This guarantees with
\begin{equation}
E_{\text{avail}}(t') \;\ge\; E_{\text{req},\,\textit{EX1}\to\textit{EX2}}+\Delta,
\label{eq:guard2}
\end{equation}
where $t'=t+T_c+T_p+T_{\text{inf,EX1}}$.
Rearranging \eqref{eq:Eavail}–\eqref{eq:guard2} yields the minimum initial voltage required to safely initiate pipeline execution with \textit{EX1}, denoted as $V_{\text{req},\,\textit{EX1}}$, as well as the minimum initial voltage required to escalate inference execution to the high-accuracy exit point, denoted as $V_{\text{req},\,\textit{EX1}\to\textit{EX2}}$. To complete the energy-aware operation, we follow the decision rule and scheduling points introduced in Sections~\ref{sec:timing-energy} and \ref{subsec:runtime_policy}.

\section{Prototype Implementation and Results}
\label{sec:prototype_sec}
\begin{figure}[!t]
  \centering
  \includegraphics[width=3in]{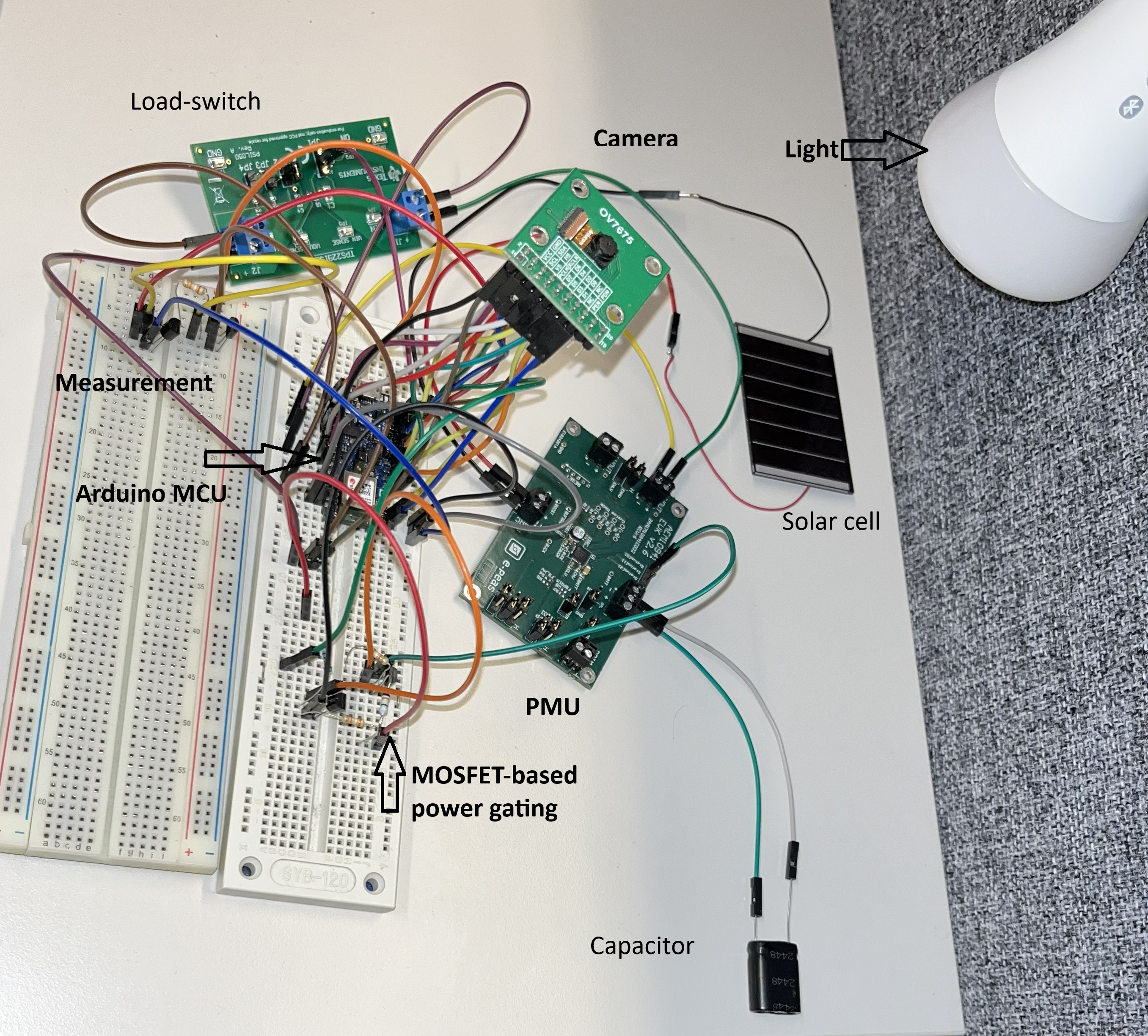}
  \caption{Experimental setup of the proposed smart-ZED showing the controllable light source, PMU, load switch, and camera interface.}
  \label{fig:prototype}
\end{figure}
In this section, we present the prototype implementation, as shown in Fig.~\ref{fig:prototype}, which is developed to validate the feasibility of the proposed approach in a real-world setting. We describe the hardware platform and the key components that enable energy-aware operation, and discuss how the implementation improves upon prior techniques with respect to the three evaluation criteria considered: energy consumption, accuracy, and memory footprint.

\subsection{Feeding and producing energy subsystem}
We utilize a controlled artificial light source in conjunction with energy harvesting hardware. Specifically, a Philips Hue White A21 LED bulb provides programmable luminous flux of up to 1600 lm, with output levels controlled via Bluetooth or Zigbee to emulate diverse lighting conditions. The incident light is converted to electrical power by a Panasonic AM\mbox{-}5608 photovoltaic module comprising six amorphous-silicon cells, whose output feeds the node’s energy-harvesting front end and downstream power-management stage. 

\subsection{Power Management Unit}
\label{sec:PMU}
\begin{figure}
  \centering
  \includegraphics[width=3in]{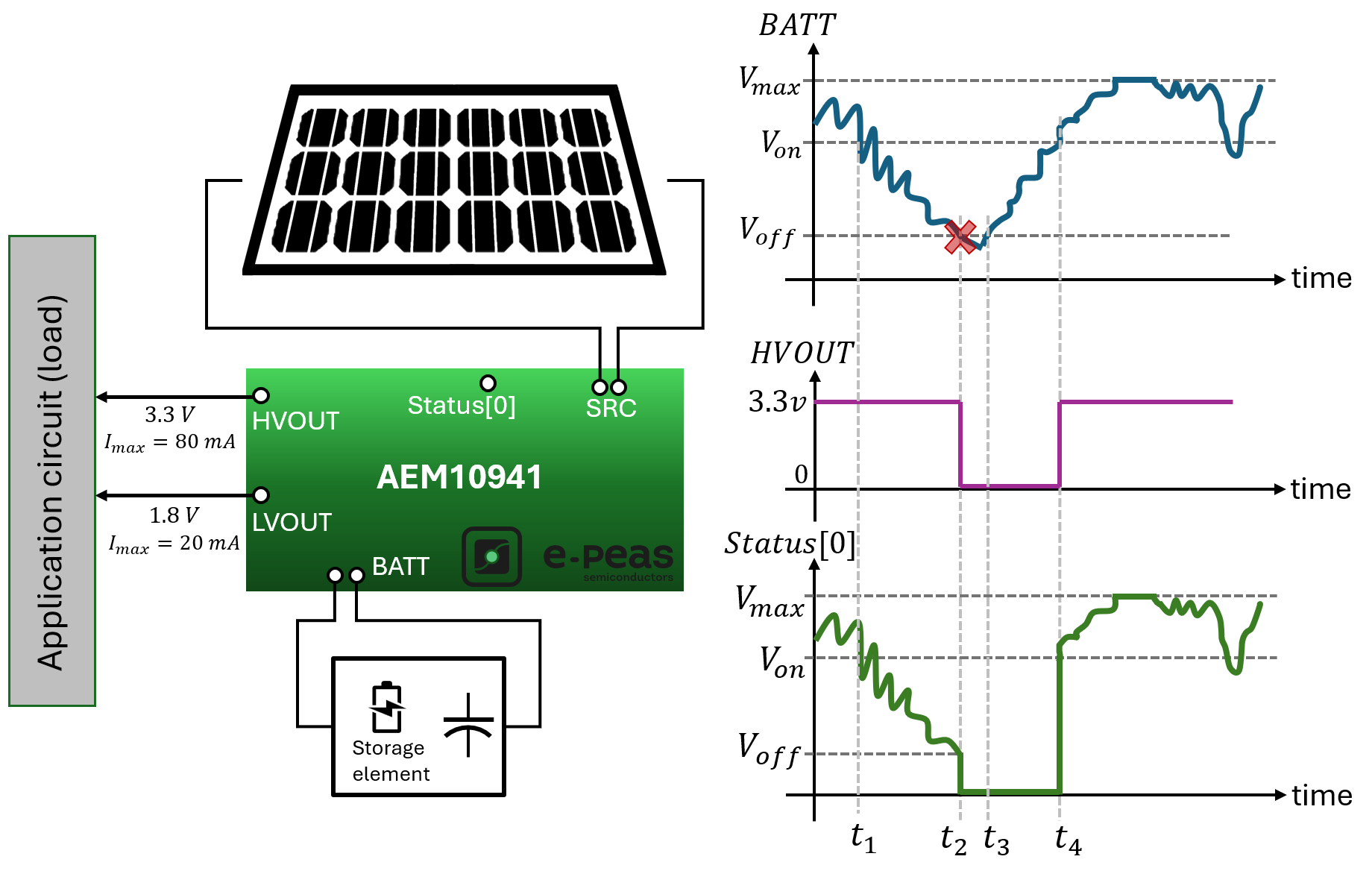}
  \caption{Voltage profile measured at the pins of the PMU.}
  \label{fig:PMU}
\end{figure}
The AEM10941 manufactured by e-peas is an integrated PMU for solar harvesters connected to the \texttt{SRC} pin. It extracts DC power from up to seven-cell solar panels and stores it in a storage capacitor. Furthermore, it regulates the stored energy to provide a stable supply for the load and peripherals. The device can power the load while charging the storage element and includes two on-chip low-dropout (LDO) regulators. One is a low-voltage output (LVOUT) selectable at $1.2\,\mathrm{V}$ or $1.8\,\mathrm{V}$, and another is a high-voltage output (HVOUT) programmable from $1.8$ to $4.1\,\mathrm{V}$. Output selections are made via the on-board configuration pins $\texttt{CFG[2{:}0]}$.
Because the load is an Arduino-class MCU with $3.3\,\mathrm{V}$-only I/O (e.g., the Nano 33 BLE Sense, nRF52840), we select the $\texttt{CFG[2{:}0]}$ setting that programs HVOUT to $3.3\,\mathrm{V}$, ensuring electrical compatibility and preventing over-voltage at the MCU pins (cf. Fig.~\ref{fig:PMU}).

The charging operation is governed by three thresholds on the storage capacitor: (i) $V_{\max}$, the maximum allowable capacitor voltage; (ii) $V_{\mathrm{on}}$, the minimum storage voltage (after a cold start) required before enabling the LDO regulators; and (iii) $V_{\mathrm{off}}$, the minimum storage voltage below which the storage element is considered depleted. As illustrated in Fig.~\ref{fig:PMU}, the board operates in four modes depending on the capacitor voltage $V_c$ at the \texttt{BATT} pin \cite{toward_energy_aware}:
\begin{enumerate}[label=(\alph*)]
  \item \textbf{$V_c = V_{\max}$:} Charging is disabled; the board supplies only the load.
  \item \textbf{$V_{\mathrm{on}} \le V_c < V_{\max}$:} When harvested power is available, the board supplies the load and simultaneously charges the storage capacitor.
  \item \textbf{$V_{\mathrm{off}} < V_c < V_{\mathrm{on}}$:} A hysteretic region with two sub-states (see $[t_1,t_2]$ and $[t_3,t_4]$ in Fig.~\ref{fig:PMU}):
    \begin{itemize}
      \item If $V_c$ has previously exceeded $V_{\mathrm{on}}$, incoming energy can both power the load and charge the storage element (cf. $[t_1,t_2]$ in Fig.~\ref{fig:PMU}).
      \item If $V_c$ has previously fallen below $V_{\mathrm{off}}$ (cold-start recovery), incoming energy is used solely to recharge the capacitor. In this case, HVOUT is held at $0\,\mathrm{V}$ (disabled) until $V_c$ reaches $V_{\mathrm{on}}$ at which point the LDOs are enabled (cf. $[t_3,t_4]$ in Fig.~\ref{fig:PMU}).
    \end{itemize}
  \item \textbf{$V_c < V_{\mathrm{off}}$:} Only the storage capacitor is charged, and the outputs remain disabled.
\end{enumerate}
A typical power failure event is indicated by a red cross in Fig.~\ref{fig:PMU}. The energy-aware strategy employed in this work is designed to prevent such occurrences.

\subsection{Microcontroller and Sensing Subsystem}
The device’s primary computation and control platform is the Arduino Nano 33 BLE Sense, which is built around the Nordic Semiconductor nRF52840 SoC. The SoC features a 64 MHz Arm Cortex-M4F processor with single-precision floating-point hardware acceleration, supporting efficient execution of signal-processing workloads and real-time control algorithms. Furthermore, the SoC includes 1 MB of embedded flash memory, which in this application is used to store the trained model and associated firmware, as well as 256 KB of SRAM dedicated to runtime operations such as data buffering, sensor handling, and stack/heap management.

To support the person-detection functionality required in the smart-ZED system, we employ the Arducam 0.3MP OV7675 camera module. The OV7675 is a compact, low-power image sensor designed for embedded vision applications where resource efficiency and real-time performance are essential. It supports different resolutions up to $640 \times 480$ pixels, which offer sufficient spatial detail for lightweight inference.

\subsection{Low-Power Design Considerations}
\label{sec:power_gating_introducing}
\begin{figure}[!t]
\centering
\includegraphics[width=2.3in]{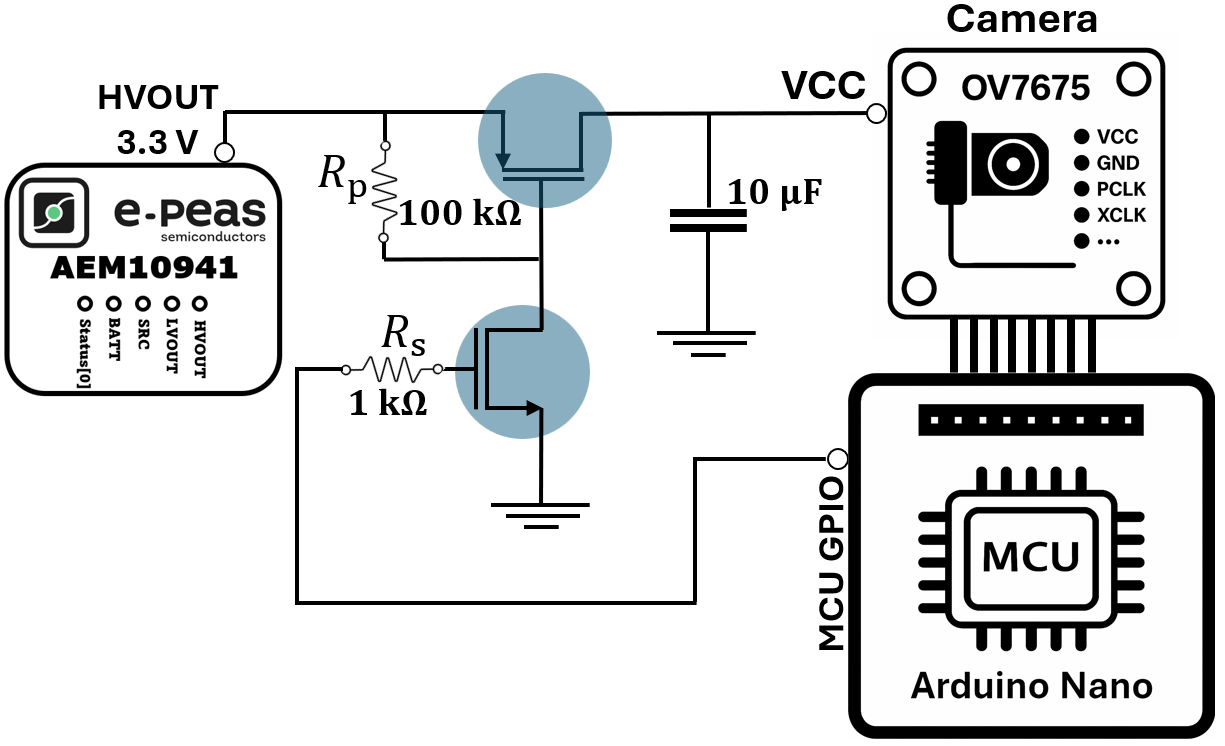}
\caption{Circuit employing a power-gating MOSFET to fully disconnect the camera module from its supply rail.}
\label{fig:power_gating}
\end{figure}
The power consumption of the camera module is approximately $98$ mW during active image capture. Although the standby power is significantly lower, keeping the sensor fully powered between frame acquisitions would considerably increase the overall energy consumption of the device. Therefore, to minimize unnecessary energy use, the system should place the camera module into standby or completely power it down between frame acquisitions.
However, powering the camera module directly from a GPIO pin would overload the pin and may prevent reliable operation or even damage the MCU. This can be seen from the fact that the module’s load current ($\approx 30$ mA, potentially higher during transients) exceeds the MCU's GPIO current limit ($\approx 15$ mA per pin).
Following this, the authors in~\cite{toward_energy_aware, model_selection} proposed an auxiliary circuit to address this issue, which is based on using a load switch controlled by the MCU's GPIO.
In this work, we propose a method that consumes less energy than a load switch that does not rely on the direct use of the pin, utilizing a MOSFET-based power-gating scheme, as shown in Fig.~\ref{fig:power_gating}

The P-MOSFET is placed between the regulated supply rail (\(\text{HVOUT}\)) and the camera's VCC pin, so that in the OFF state the camera is fully disconnected from the supply and its leakage current is limited to the MOSFET's off-state leakage. A P-MOSFET is selected because it is well-suited for high-side switching, providing a straightforward and reliable means of controlling the supply line. In contrast, a low-side N-MOSFET configuration can also be implemented where the camera’s ground return is switched between the module and the system's ground. However, this approach is often less desirable, since many camera modules depend on a stable and uninterrupted ground reference to maintain proper operation.

The power-gating switch is controlled by an MCU’s GPIO. To obtain an active-HIGH control signal, the gate of the P-MOSFET is driven via an N-MOSFET. When the GPIO output is HIGH, the N-MOSFET pulls the P-MOSFET gate low, resulting in a sufficiently negative gate--source voltage to turn the P-MOSFET fully on and connect \(\text{HVOUT}\) to the camera VCC pin. In contrast, when the GPIO output is LOW, the N-MOSFET switches off and a pull-up resistor (i.e., $R_\text{p}$) returns the P-MOSFET gate to \(\text{HVOUT}\), turning the P-MOSFET off and removing the supply from the camera. It is worth noting that a small series resistor (e.g., $R_\text{s}=1 k\Omega$) is included to limit the instantaneous current spikes drawn from the GPIO pin during switching. Furthermore, to help stabilize the supply seen by the camera and reduce voltage dips when the camera current suddenly changes, we employ a capacitor $10\mu\text{F}$.

The selected P-MOSFET has a low on-resistance \(R_{\text{DS(on)}}\) (on the order of a few tens of milliohms) at \(V_{\text{GS}} \approx -3.3\,\text{V}\). At the camera operating current of \(\sim 30\,\text{mA}\), this results in a voltage drop of only a few millivolts and a conduction loss below \(0.1\,\text{mW}\), both negligible compared to the camera's own power consumption. Consequently, the voltage seen at the camera's VCC pin is effectively equal to the system supply, and the MOSFET current rating provides ample margin over the required camera current. This architecture enables aggressive power-gating of the camera with minimal component overhead while preserving simple, active-HIGH digital control from the MCU. It is worth noting that, since the absolute maximum rating for the supply voltage is \(4.5\,\text{V}\), the chosen \(3.3\,\text{V}\) rail remains well below the damage threshold, although it slightly exceeds the recommended operating maximum of \(3.0\,\text{V}\).

\subsection{Energy-Aware Management Subsystems}
\label{subsec:energy_aware_measurement}
\begin{figure}[!t]
\centering
\includegraphics[width=3in]{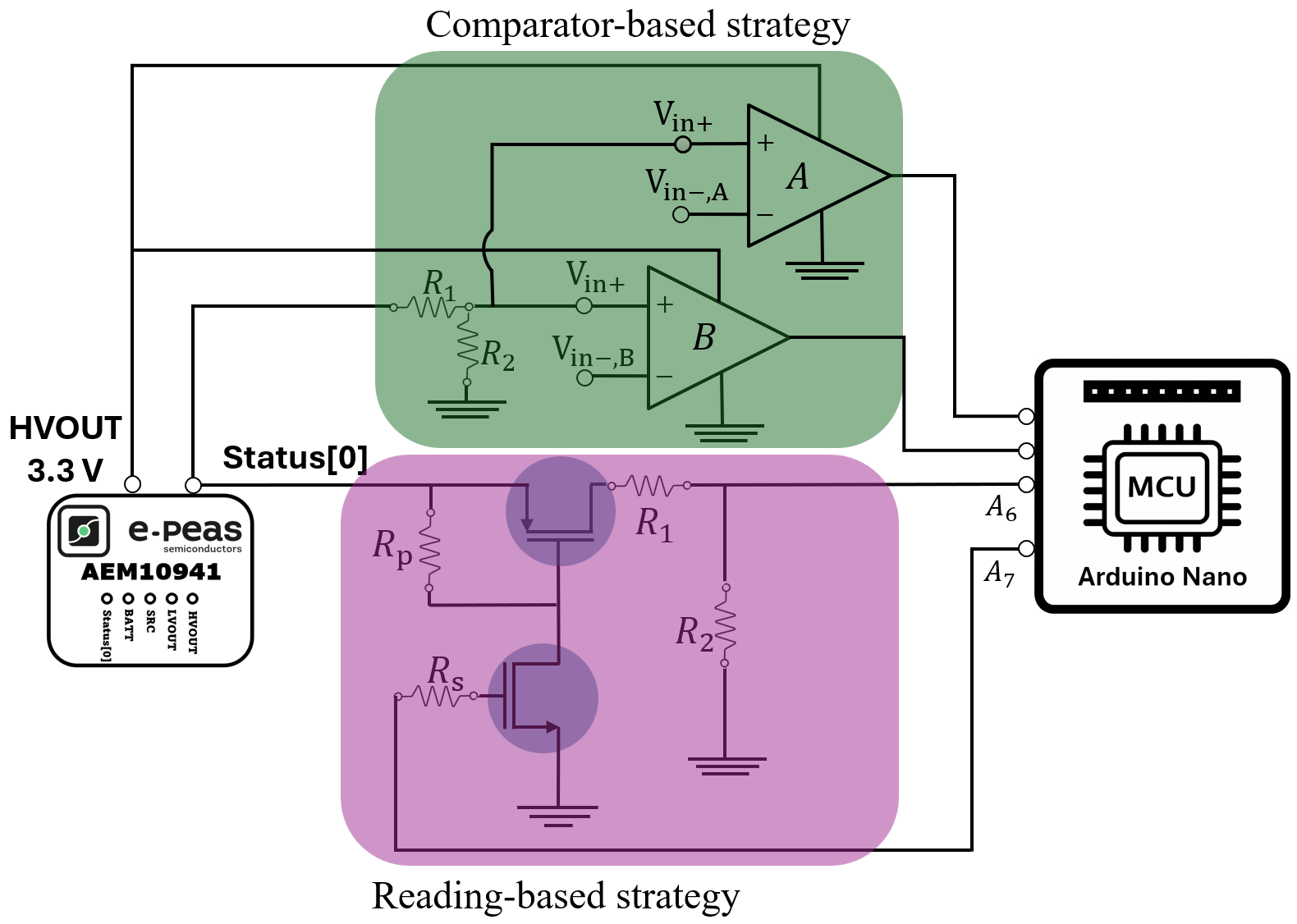}
\caption{Two divider-based strategies for monitoring available energy to support energy-aware device operation.}
\label{fig:divider}
\end{figure}
As shown in Fig.~\ref{fig:different_states}, operating the device in an energy-aware manner requires measuring the available energy at least once within each window. To this end, we consider two measurement strategies that operate at \texttt{Status[0]} pin, one of which can be selected to enable energy-aware operation. The first relies on hardware comparators to check whether the supply voltage exceeds the required thresholds (i.e., $V_{\text{req},\,\textit{EX1}}$ or $V_{\text{req},\,\textit{EX1}\to\textit{EX2}}$), while the second uses an ADC channel of the MCU to read the voltage directly. Before examining these strategies in detail, it is important to note that the reference voltage may reach up to $4.5\,\text{V}$ (cf. Section~\ref{sec:PMU}), exceeding the allowable $3.3\,\text{V}$ input threshold of the I/O GPIO pins. To ensure the voltage remains within a safe range for the circuit, a resistive divider is employed to scale the value of $V_{\text{c}}$ accordingly.

In the comparator-based approach, illustrated in Fig.~\ref{fig:divider}, we embed two comparators in the device to assess the available energy either at \(s_k\) or at a potential low-confidence point after \textit{EX1}. Specifically, the threshold of comparator~A is set to the minimum required voltage defined as $V_{\text{req},\,\textit{EX1}}$, and the threshold of comparator~B is set to the value defined as $V_{\text{req},\,\textit{EX1}\to\textit{EX2}}$. With the divider composed of \(R_1\) and \(R_2\), the voltage at the non-inverting input is $V_{\text{in}+} = \alpha_0 V_{\text{c}}$, where $\alpha_0=\frac{R_2}{R_1 + R_2}$. Accordingly, the reference thresholds applied to the inverting inputs are chosen as
\[
V_{\text{in-},A} = \alpha_0 V_{\text{req},\,\textit{EX1}} \quad \text{and} \quad
V_{\text{in-},B} = \alpha_0 V_{\text{req},\,\textit{EX1}\to\textit{EX2}},
\]
for comparators~A and~B, respectively. In this configuration, each comparator drives a dedicated digital GPIO of the MCU, which is asserted HIGH when the capacitor voltage reaches the corresponding minimum required value.

In the reading-based strategy, the device measures the capacitor voltage at \texttt{Status[0]} pin using an ADC channel (i.e., $A_6$) on the nRF52840, configured as a high-impedance input. The resistive divider is inserted before the ADC GPIO pin to squash the capacitor’s measured voltage to a level that remains within the ADC’s safe operating range. Therefore, the voltage at $A_6$ would be $V_{A_6}= \alpha_0 V_{\text{Status[0]}}$.
Additionally, to avoid a continuous leakage path through this divider, an MCU's GPIO (i.e., $A_7$) controls a series MOSFET, effectively duty-cycling the measurement path (cf. Fig.~\ref{fig:divider}). Consequently, the divider is only connected during brief sampling intervals and is otherwise disconnected, minimizing quiescent energy loss.

\subsection{Current Consumption Profiling}
\label{sec:current_cons}
\begin{figure*}[!t]
    \centering
    \begin{minipage}{0.24\textwidth}
        \centering
        \includegraphics[width=\linewidth]{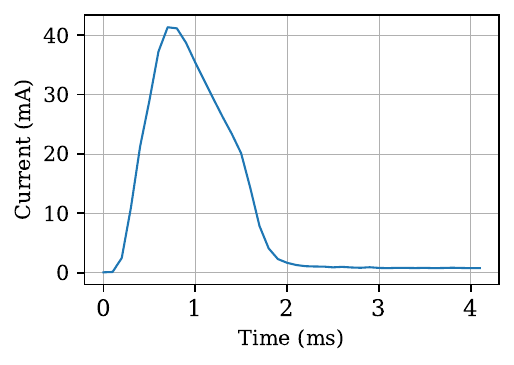}
        \\[-2pt] 
        (a)
    \end{minipage}
    \begin{minipage}{0.24\textwidth}
        \centering
        \includegraphics[width=\linewidth]{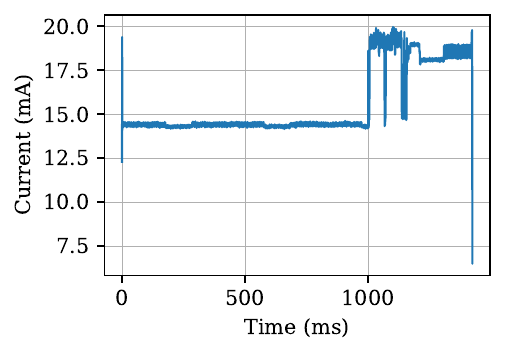}
        \\[-2pt]
        (b)
    \end{minipage}
    \begin{minipage}{0.24\textwidth}
        \centering
        \includegraphics[width=\linewidth]{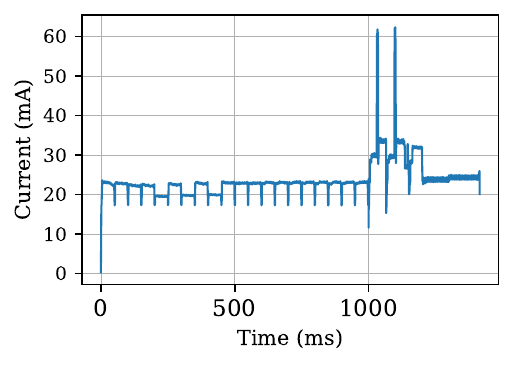}
        \\[-2pt]
        (c)
    \end{minipage} 
    \begin{minipage}{0.24\textwidth}
        \centering
        \includegraphics[width=\linewidth]{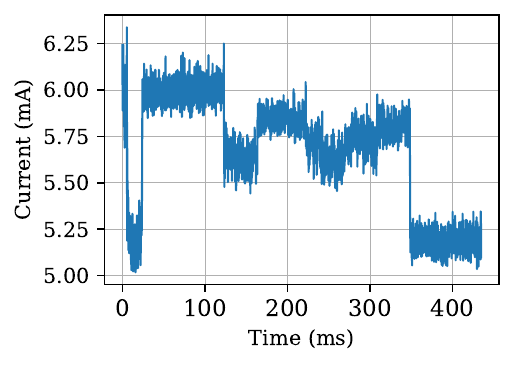}
        \\[-2pt]
        (d)
    \end{minipage}
    \begin{minipage}{0.25\textwidth}
        \centering
        \includegraphics[width=\linewidth]{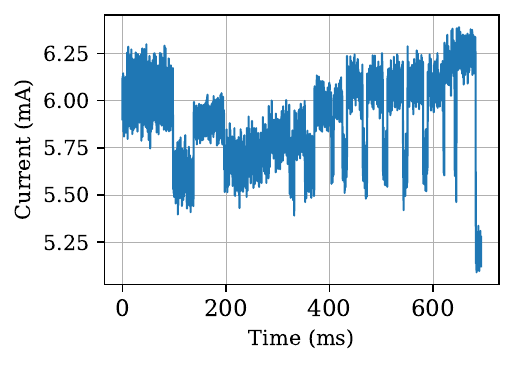}
        \\[-2pt]
        (e)
    \end{minipage}
    \begin{minipage}{0.25\textwidth}
        \centering
        \includegraphics[width=\linewidth]{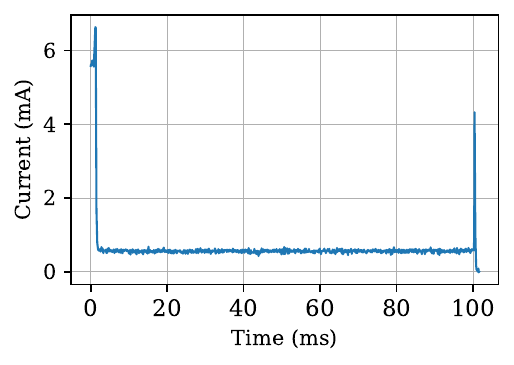}
        \\[-2pt]
        (f)
    \end{minipage}
    \begin{minipage}{0.25\textwidth}
        \centering
        \includegraphics[width=\linewidth]{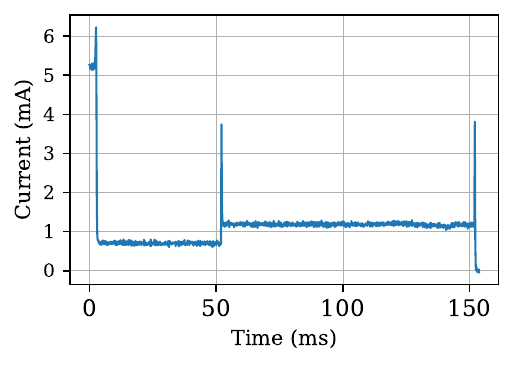}
        \\[-2pt]
        (g)
    \end{minipage}
    \caption{Current consumption associated with each operation in the processing pipeline: (a) measurement of the capacitor voltage; (b) image capture and pre-processing controlled by the MOSFET power-gating; (c) image capture and pre-processing supplying with a load switch; (d) inference using \textit{EX1}; (e) inference using \textit{EX2}; (f) activation of the LED blue channel; and (g) activation of the LED green and red channels.}
\label{fig:current_consumption}
\end{figure*}
\begin{table*}
    \caption{Current consumption for different states.}
    \centering
    \begin{tabular}{lllll}\hline
         \textbf{State}    &   \textbf{Current} (mA)   &  \textbf{Execution time} (ms)  &  \textbf{Energy consumption} (mJ)  \\\hline
         \makecell[l]{Capturing and pre-processing \\ (MOSFET-based)}& $15.5868\pm 1.9482$ & $1417.2\pm2.52$ & $72.896$  \\
         \makecell[l]{Capturing and pre-processing \\ (load switch-based \cite{toward_energy_aware})} & $23.7491\pm 4.4738$ & $1409.2\pm2.74$ & $110.442$  \\
         Inference via \textit{EX1}   & $5.6646 \pm 0.3026$  & $434.1\pm 2.05$ & $8.118$  \\
         Inference via \textit{EX2}   & $5.8884 \pm 0.2251$  & $689.1\pm 3.54$ & $13.390$  \\
         Measuring the voltage        & $10.418 \pm 14.17$   & $4.145\pm0.97$  & $0.8934$  \\
         LED (green diode)            & $0.7162 \pm 0.0312$  & $50$ & $0.1182$ \\
         LED (blue diode)             & $0.5714 \pm 0.0257$  & $100$ & $0.1885$ \\
         LED (red diode)              & $1.1912 \pm 0.0431$  & $100$ & $0.3931$ \\ \hline
    \end{tabular}
    \label{tab:energy_consumption_numbers}
\end{table*}

In this subsection, we analyze the current consumption at each stage of the processing pipeline and compare the results across the proposed approaches, as illustrated in Fig.~\ref{fig:current_consumption}. Current measurements are obtained using the Power Profiler Kit II from Nordic Semiconductor. Fig.~\ref{fig:current_consumption}(a) presents the current profile during the measurement phase, which must be completed within the permissible time window and may occur at the end of the inference performed by \textit{EX1} when a low-confidence outcome is produced (i.e., $O_1 \in {(\gamma_1,\gamma_2)}$). Fig.~\ref{fig:current_consumption}(b) and (c) then report the current consumption for steps P1 and P2 of the pipeline under two different power-gating strategies. Specifically, Fig.~\ref{fig:current_consumption}(b) shows the results obtained using the MOSFET-based power-gating strategy introduced in Section~\ref{sec:power_gating_introducing}, whereas Fig.~\ref{fig:current_consumption}(c) illustrates the current consumption when a load switch is used to supply power to the camera through one of the MCU’s GPIO pins, following the approach adopted in work~\cite{toward_energy_aware}. The results indicate that the load switch–based method leads to approximately $34\%$ higher energy consumption compared to our proposed approach.

The current consumption during the inference phase concluding at \textit{EX1} and \textit{EX2} is reported in Fig.~\ref{fig:current_consumption}(d) and (e), respectively. As shown, inference performed at \textit{EX1} requires an execution time that is reduced by a factor of $1.58$ compared to \textit{EX2}.

Finally, Fig.~\ref{fig:current_consumption}(f) and (g) show the measured supply current as a function of time during LED activation. In Fig.~\ref{fig:current_consumption}(f), only the blue channel of the MCU’s on-board RGB LED is enabled, whereas Fig.~\ref{fig:current_consumption}(g) corresponds to the case where the device completes the inference via \textit{EX2}, indicated by turning on the green and red channels, respectively. As observed, the current drawn when each color is active is different. This can be seen from the fact that per-color LED current is not identical because the red/green/blue dies have different forward voltages despite using the same series resistor.
Table~\ref{tab:energy_consumption_numbers} reports the mean and standard deviation of the current consumption at each step. In addition, using \eqref{eq:Et}, the corresponding energy consumption values are also provided.

\subsection{Performance Improvement of the Proposed Approach}
\label{sec:tradeoff}
\begin{table}
\centering
\caption{Experimental setup.}
\label{tab:experimental_setup}
\begin{tabular}{lll}\hline
\textbf{Parameter} & \textbf{Symbol}  & \textbf{Value} \\\hline
Window duration    & $T$              & $10$s \\
Capacitance        & $C$              & $1.5$F \\
Deadline           & $\tau$           & $4$s \\
Number of measurements for adaptive rule   & $N$ & $20$ \\
Minimum voltage    & $V_{\text{off}}$ & $3.6$v \\
Maximum voltage    & $V_{\text{max}}$ & $4.5$v \\ 
Turn-on voltage    & $V_{\text{on}}$  & $3.92$v \\
Low threshold      & $\gamma_1$       & $0.3$\\
High threshold     & $\gamma_2$       & $0.7$\\\hline
\end{tabular}
\end{table}
\begin{figure}
  \centering
  \includegraphics[width=3.4in]{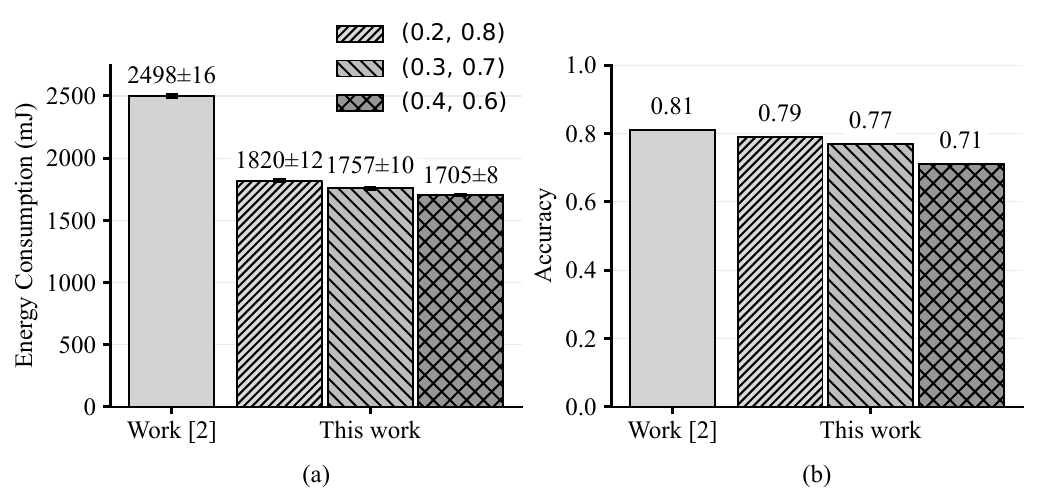}
  \caption{Performance comparison between this work for different values for $(\gamma_1,\gamma_2)$ and Work\cite{toward_energy_aware} in terms of (a) energy consumption over a $200$-s operation period and (b) accuracy.}
  \label{fig:energy_accuracy_comparison}
\end{figure}
\begin{figure}
  \centering
  \includegraphics[width=3in]{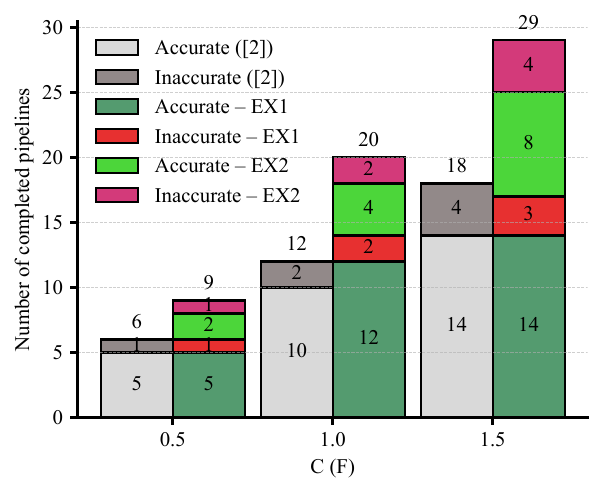}
  \caption{Performance comparison between the proposed method and \cite{toward_energy_aware} for different values of $C$, in terms of the number of completed pipelines, under a fixed harvested current of $I_h=2$ mA.}
  \label{fig:completed_pipeline_comparison}
\end{figure}

Here, we evaluate the performance improvements of the proposed MOSFET-based approach by comparing it with the method presented in~\cite{toward_energy_aware} also for person detection. The experimental setup for this comparison is listed in Table~\ref{tab:experimental_setup}.
To illustrate the performance gains, Fig.~\ref{fig:energy_accuracy_comparison} compares the energy consumption and accuracy under different levels of $(\gamma_1,\gamma_2)$. These metrics are evaluated over a $200$-s device operation period, assuming an initial capacitor voltage of $4.5$~V. To ensure reliable results, each experiment was performed $10$ times, and the reported result corresponds to the average of the obtained outcomes. As shown in Fig.~\ref{fig:energy_accuracy_comparison}, the proposed approach achieves a substantial reduction in energy consumption (i.e., around $29.6\%$ for $(\gamma_1,\gamma_2)=(0.3,0.7)$), albeit at the cost of a slight decrease in accuracy. Furthermore, it can be concluded that narrowing the ambiguity band leads to reduced energy consumption and decreased accuracy. This occurs because a larger number of inputs perform inference at \textit{EX1}.
\begin{figure}
    \centering
    \begin{minipage}{0.5\textwidth}
        \centering
        \includegraphics[width=\linewidth]{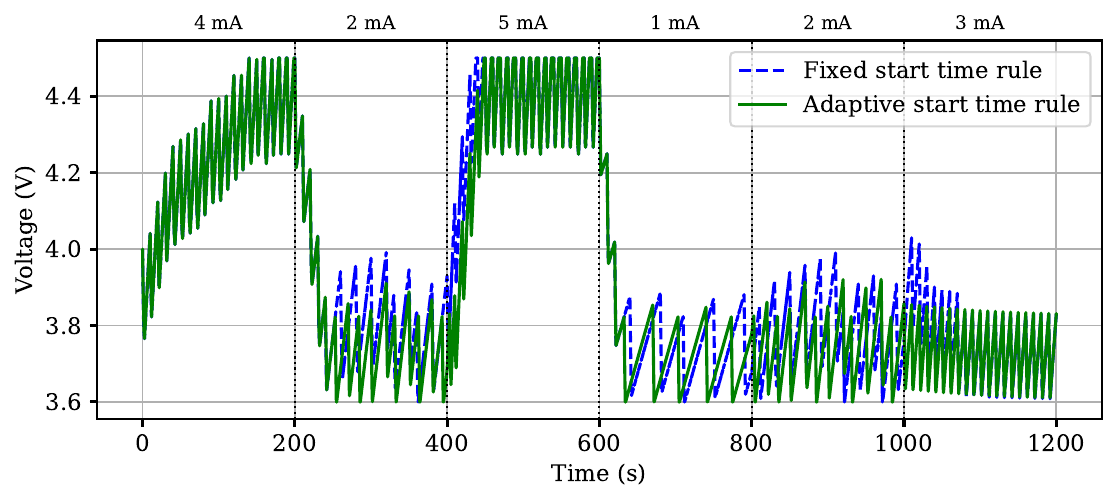}
        \\[-2pt]
    \end{minipage}
    \begin{minipage}{0.3\textwidth}
        \centering
        \includegraphics[width=\linewidth]{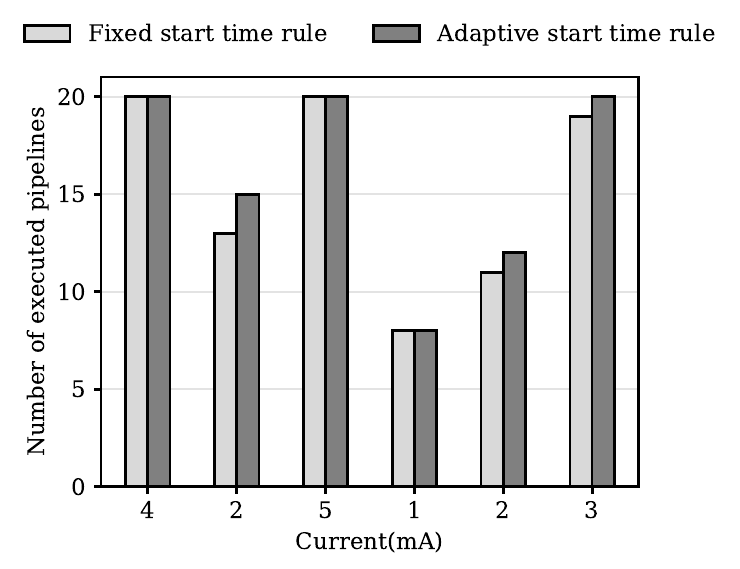}
        \\[-2pt]
    \end{minipage} 
    \caption{Capacitor voltage trajectories under piecewise-constant harvested current levels, each applied for a $200$-s interval, for fixed and adaptive start time rules (top). Corresponding comparison of completed pipeline counts per interval (bottom), shown as the difference between the totals achieved by the two rules.}
\label{fig:execution_consumption}
\end{figure}
To further highlight the improvement achieved by the proposed approach, we evaluate the number of completed pipelines before power depletion and the effective engagement of the energy-aware strategy. In this experiment, the initial voltage across the capacitor is set to $4$ V, and the harvested current is fixed at $I_h=2$ mA. Fig.~\ref{fig:completed_pipeline_comparison} reports the number of completed pipelines for different values of the capacitance $C$. As observed, the proposed approach consistently results in a higher number of completed pipelines and successful inferences compared to~\cite{toward_energy_aware}, demonstrating its effectiveness under identical energy-harvesting conditions.

\subsection{Non-constant Harvested Current}
\label{subsec:policies}
To evaluate the differences between the execution policies introduced in Section~\ref{sec:timing-energy} (see Fig.~\ref{fig:execution_rules}), we conduct experiments using a low-capacity capacitor with $C=0.1$~F. The capacitor is initially charged to $4$~V, and the harvested current is modeled as a piecewise-constant signal with $200$-s intervals. Fig.~\ref{fig:execution_consumption} (top) reports the capacitor voltage over time under both policies for different harvested-current levels. The results show that when the harvested current exceeds $4$~mA, the harvested energy exceeds the energy consumption, causing the capacitor voltage to increase and eventually saturate at $V_{\max}$. Under this condition, the device does not experience a power failure. In addition, Fig.~\ref{fig:execution_consumption} (bottom) highlights the advantage of the adaptive start time rule, which utilizes the available energy more effectively. Specifically, the rule evaluates feasibility throughout the window rather than only at its start, reducing missed opportunities to execute the pipeline when sufficient energy becomes available later within the same window.

\section{Conclusion}
\label{sec:conclu}
In this paper, we presented an energy-aware multi-exit TinyML architecture for a ZED that explicitly utilizes for the trade-off among memory footprint, energy consumption, and inference accuracy. To demonstrate feasibility, we implemented a person-detection application and validated deployment on a hardware prototype capable of running the required on-device intelligence under energy constraints.
We first reviewed the multi-exit ML model and trained it on the COCO dataset, systematically analyzing how accuracy and the number of produced inferences at different exit points vary with key hyperparameters. To cope with intermittent energy availability and to keep the device in an energy-efficient operating regime, we proposed a control policy called a purely energy-aware confidence-gated policy that leverages model certainty to regulate when inference is executed.
We then introduced a MOSFET-based circuit for a control strategy to disconnect unused modules during idle periods. Experimental results showed that this mechanism reduces energy consumption by approximately $34\%$ related to the load switch-based approach for pipeline steps P1 and P2.
In addition, we proposed and evaluated two strategies for monitoring available energy and triggering appropriate energy-aware fashion. Notably, the overhead of the energy-measurement operation itself was found to be negligible relative to the overall energy consumption.
Finally, we benchmarked the proposed energy-aware multi-exit TinyML framework against the single-exit TinyML approach and observed consistent improvements in energy–accuracy behavior. In particular, the proposed design enables a higher number of inferences with improved cumulative accuracy while consuming less energy, supporting the practicality of energy-aware multi-exit TinyML for smart-ZEDs deployments under intermittent power.


%

\ifCLASSOPTIONcaptionsoff
  \newpage
\fi



%
\bibliographystyle{IEEEtran}
\bibliography{reference.bib}

%








\end{document}